\documentclass[aps,superscriptaddress,showpacs,floatfix,nobibnotes]{revtex4}

\usepackage{subfigure}
\usepackage{epsfig}
\usepackage{epstopdf}
\usepackage{color}
\usepackage{graphicx}
\usepackage{longtable}
\usepackage{CJK}
\usepackage{amsmath,bbold}
\usepackage{float}
\newcommand{\blue}[1]{{ \color{blue} #1 }}
\ifx\pdfoutput\undefined
\usepackage[dvips]{graphicx}
\usepackage[dvipdfm,
  pdfstartview=FitH,
     CJKbookmarks=true,
          bookmarksnumbered=true,
          bookmarksopen=true,
         colorlinks=true,
          pdfborder=002,
          citecolor=blue%
           ]{hyperref}
\AtBeginDvi{} 
\else
\usepackage[
          pdfstartview=FitH,
          CJKbookmarks=true,
          bookmarksnumbered=true,
           bookmarksopen=ture,
           colorlinks=true,
           citecolor=blue
           ]{hyperref}
\usepackage{graphicx}
\usepackage{longtable}

\begin{document}

\title{Construction of Basis Vectors For Symmetric \\Irreducible Representations of $O(5)\supset O(3)$}

\author{Feng Pan}
\affiliation{Department of Physics, Liaoning Normal University,
Dalian 116029, China}\affiliation{Department of Physics and
Astronomy, Louisiana State University, Baton Rouge, LA 70803-4001,
USA}\affiliation{School of Mathematics and Physics, The University of Queensland,
Brisbane, Qld 4072, Australia}

\author{Lina Bao}
\affiliation{Department of Physics, Liaoning Normal University,
Dalian 116029, China}

\author{Yao-Zhong Zhang}
\affiliation{School of Mathematics and Physics, The University of Queensland,
Brisbane, Qld 4072, Australia}

\author{Jerry P. Draayer}
\affiliation{Department of Physics and Astronomy, Louisiana State
University, Baton Rouge, LA 70803-4001, USA}
\date{\today}

\begin{abstract}
A recursive method for construction of symmetric irreducible representations
of $O(2l+1)$ in the $O(2l+1)\supset O(3)$ basis for identical boson systems is proposed.
The formalism is realized based on the group chain $U(2l+1)\supset U(2l-1)\otimes U(2)$,
of which the symmetric irreducible representations are simply reducible.
The basis vectors of the $O(2l+1)\supset O(2l-1)\otimes U(1)$ can easily be constructed
from those of $U(2l+1)\supset  U(2l-1)\otimes U(2)\supset O(2l-1)\otimes U(1)$
with no boson pairs, from
which one can construct symmetric irreducible representations of $O(2l+1)$ in the $O(2l-1)\otimes U(1)$
basis when all symmetric irreducible representations of $O(2l-1)$ are known. As a starting point,
basis vectors of symmetric irreducible representations of $O(5)$ are constructed in the $O_{1}(3)\otimes U(1)$ basis.
Matrix representations of  $O(5)\supset O_{1}(3)\otimes U(1)$, together with
the elementary Wigner coefficients, are presented.
After the angular momentum projection, a three-term relation in determining the expansion
coefficients of the $O(5)\supset O(3)$ basis vectors in terms of those of
the $O_{1}(3)\otimes U(1)$ is derived. The eigenvectors
of the projection matrix with zero eigenvalues constructed according to the three-term relation
completely determine the  basis vectors of
$O(5)\supset O(3)$. Formulae for evaluating the
elementary Wigner coefficients of $O(5)\supset O(3)$ are derived explicitly.
Analytical expressions
of some elementary Wigner coefficients of $O(5)\supset O(3)$
for the coupling $(\tau~0)\otimes (1~0)$ with resultant
angular momentum quantum number $L=2\tau+2-k$ for
$k=0,2,3,\cdots,6$ with a multiplicity $2$ case for $k=6$ are
presented.
\end{abstract}

\pacs{03.65.Fd, 02.20.Qs, 21.60.Fw, 02.30.Ik}

\maketitle

\section{introduction}
The orthogonal group $O(2l+1)$ and its Lie algebra occurs naturally in the classification of
many-particle states of identical bosons with angular momentum $l$
referred to as the $l$-bosons hereafter under the group chain
$U(2l+1)\supset O(2l+1)\supset O(3)$,
which is useful in atomic, molecular, and nuclear physics \cite{ham,judd, wybourne}.
However, to construct $U(2l+1)\supset O(2l+1)\supset O(3)$
basis vectors is not easy mainly
because the missing label problem in the reduction $O(2l+1)\downarrow O(3)$,
which is not multiplicity-free in general.
 {A non-trivial simplest} case is to construct symmetric irreducible representations  {(irreps)}
of the $O(5)$ group in the $O(3)$ basis for identical $d$-bosons useful
in the nuclear collective model~\cite{bohr,rowe} and the interacting boson model for nuclei~\cite{Iac}.
Because of  {its} physical importance, there  {have been a lot of attempts to construct} the $O(5)\supset O(3)$ basis vectors~\cite{will,corr,bud,cha1,cha2,cas,ghe,ash,ber,roh,szp,hess,sei,van1,van2,yan}.
Most notably, Rowe, Hecht, and many others in a series of papers~\cite{rowe941,rowe942,rh, tur} established
the vector-coherent-state (VCS) representations of $O(5)\supset O(3)$ and constructed
the $O(5)$ spherical harmonics~\cite{rt}.
As shown in \cite{rt}, the $O(5)$ spherical harmonics
are useful for calculating the Wigner coefficients
of $O(5)\supset O(3)$, which can also be computed in  {a number of other ways,} for example, those shown in ~\cite{sun,cap1}.

There are many subgroup chains of $O(5)$, for example, those  shown in \cite{cap2}.
Besides the VCS construction, the basis vectors of $O(5)\supset O(3)$ can
be expanded in terms of any one of other group chains of $O(5)$.
Similarly, for identical $l$-boson systems, basis vectors of $O(2l+1)\supset O(3)$
may be expanded in terms of
those of $O(2l+1)\supset O(2l-1)\otimes U(1)$, which thus provides a systematic
recursive procedure to construct the basis vectors of $O(2l+1)\supset O(3)$
starting with $l=2$. In this paper, we focus on the $l=2$ case to show how the
procedure works.

\section{The $U(2l+1)\supset U(2l-1)\otimes U(2)$ basis for $l$-bosons}

Let $b^{\dagger}_{\mu}$ ($b_{\mu}$) ($\mu=-l,-l+1,\cdots,l$) be boson creation (annihilation) operators
satisfying the following commutation relations:
$$[b_{\mu},~b_{\mu^{\prime}}]=[b^{\dagger}_{\mu},~b^{\dagger}_{\mu^{\prime}}]=0,~~
[b_{\mu},~b^{\dagger}_{\mu^{\prime}}]=\delta_{\mu\mu^{\prime}}.\eqno(1)$$
 {The} $(2l+1)^2$ bilinear forms $\{b^{\dagger}_{\mu}b_{\mu^{\prime}}\}$
or the equivalent $O(3)$ tensors $\left(b^{\dagger}\times \tilde{b}\right)^{(k)}_{\mu}$
with $k=0,~1,\cdots,2l$ and $\mu=k,k-1,\cdots,-k$ for fixed $k$, in which $\tilde{b}_{\mu}=(-)^{l-\mu}b_{-\mu}$,
generate the $U(2l+1)$ algebra, where  {for convenience the Lie group notation
is also used to denote the corresponding Lie algebra.}
It is well known that $(2l+1)l$ operators $\left(b^{\dagger}\times
\tilde{b}\right)^{(k)}_{\mu}$  with $k={\rm odd}$
generate the subalgebra $O(2l+1)$. Moreover, $\left(b^{\dagger}\times \tilde{b}\right)^{(1)}_{\mu}$
are generators of the $O(3)$ subalgebra. In many physics applications,
one needs to construct the $U(2l+1)$ basis adapted to the group chain
$U(2l+1)\supset O(2l+1)\supset O(3)$. The reduction of  $O(2l+1)\downarrow O(3)$
is not multiplicity-free except the trivial $l=1$ case.

\vskip .3cm
Actually, there is a simple mathematical basis  {for} $U(2l+1)$ when its Lie algebra is realized
in terms of boson creation and annihilation operators.
 {The} $\left(b^{\dagger}\times \tilde{b}\right)^{(k)}_{\mu}$
with $k=1,\cdots,2l-2$ constructed from $b^{\dagger}_{\mu}$ ($b_{\mu}$) with
$\mu=-(l-1),-(l-1)+1,\cdots,l-1$ generate the $U(2l-1)$ subalgebra,
while $J_{+}=b^{\dagger}_{l}b_{-l}$, $J_{-}=b^{\dagger}_{-l}b_{l}$,
and $J_{0}={{1}\over{2}}( b^{\dagger}_{l}b_{l}-b^{\dagger}_{-l}b_{-l})$
generate the $U(2)$ subalgebra. Obviously,    $U(2l-1)\otimes U(2)$
is a subgroup of $U(2l+1)$. For a  {given irrep}
$[n\dot{0}]$ of $U(2l+1)$, the reduction
$U(2l+1)\downarrow U(2l-1)\otimes U(2)$ is simple with

$$
\begin{array}{l}
U(2l+1)~~\downarrow~~~~~~~~~~~~U(2l-1)\otimes U(2)\cr
~~~~[n\dot{0}]~~~~\downarrow~~~~\oplus^{n}_{2J=0}[n-2J~\dot{0}]\otimes J
\end{array},\eqno(2)$$
where  {for simplicity we use the spinor quantum number  $J$
to label irreps of the $U(2)$, with the corresponding basis vectors denoted as}

$$\left\vert
\begin{array}{l}
~~~~~~~~~~[n\dot{0}]\cr
~[n-2J~\dot{0}]~~~J\cr
~~~~~~~(\nu)~~~~~~m_{J}\\
\end{array}\right\rangle\equiv
\left\vert
\begin{array}{l}
~[n-2J~\dot{0}]\\
~~~~(\nu)\\
\end{array}
\right.
\left.
\begin{array}{l}
~J\\
m_{J}\\
\end{array}\right\rangle,\eqno(3)
$$
where $(\nu)$ stands for a set of quantum numbers needed to label
the irrep $[n-2J~\dot{0}]$ of $U(2l-1)$.

Then, the basis vectors
of $U(2l+1)\supset O(2l+1)\supset O(3)$ can be expanded
in terms of those of $U(2l+1)\supset U(2l-1)\otimes U(2)$ as

$$\left\vert
\begin{array}{l}
~[n\dot{0}]\cr
~(\tau\dot{0})\cr
\alpha L M_{L}\\
\end{array}\right\rangle=
\sum_{(\nu)Jm_{J}}a_{n\tau}^{(\nu)Jm_{J}}\left\vert
\begin{array}{l}
~[n-2J~\dot{0}]\\
~~~~(\nu)\\
\end{array}
\right.
\left.
\begin{array}{l}
~J\\
m_{J}\\
\end{array}\right\rangle,\eqno(4)$$
where $\tau$ is the seniority quantum
number  {for labeling the $O(2l+1)$ irrep,}
$\alpha$ is the multiplicity
label needed to distinguish from
basis vectors with
the same angular momentum $L$,
and $a_{n\tau}^{(\nu)Jm_{J}}$
is the corresponding expansion
coefficient.
We always assume that the basis vectors
of $U(2l+1)\supset U(2l-1)\otimes U(2)$ are orthonormal.

In the construction
of (4), the $l$-boson pairing operator
defined as

$$P^{\dagger}_{l}=\sqrt{1\over{2}}\sum^{l}_{\mu=-\l}(-)^{l-\mu}b^{\dagger}_{\mu}b^{\dagger}_{-\mu}\eqno(5)$$
is  {a useful construction that satisfies the following commutation relation}

$$[P_{l},~P^{\dagger\xi}_{l}]=\xi P^{\dagger\xi-1}_{l}\left(2\sum_{\mu=-l}^{l}b^{\dagger}_{\mu}b_{\mu}+2\xi+2l-1\right).\eqno(6)$$
The basis vectors of $U(2l+1)\supset O(2l+1)\supset O(3)$ with $n>\tau$ can be expressed by
those with $n=\tau$ and the pairing operator (5) as~\cite{sun,pan03}

$$\left\vert
\begin{array}{l}
~[n\dot{0}]\cr
~(\tau\dot{0})\cr
\alpha L M_{L}\\
\end{array}\right\rangle =\left[
{(2\tau+2l-1)!!\over{\xi!(2\tau+2\xi+2l-1)!!}}\right]^{1\over{2}}
P_{l}^{\dagger\xi}\left\vert
\begin{array}{l}
~[\tau\dot{0}]\cr
~(\tau\dot{0})\cr
\alpha L M_{L}\\
\end{array}\right\rangle =$$$$
\left[
{(2\tau+2l-1)!!\over{\xi!(2\tau+2\xi+2l-1)!!}}\right]^{1\over{2}}
\sum_{(\nu)Jm_{J}}a_{n\tau}^{(\nu)Jm_{J}}P_{l}^{\dagger\xi}\left\vert
\begin{array}{l}
~[\tau-2J~\dot{0}]\\
~~~~(\nu)\\
\end{array}
\right.
\left.
\begin{array}{l}
~J\\
m_{J}\\
\end{array}\right\rangle,\eqno(7)$$
where $n=\tau+2\xi$,
$\left\vert\begin{array}{l}
~[\tau\dot{0}]\cr
~(\tau\dot{0})\cr
\alpha LM_{L}\\
\end{array}
\right\rangle$
is the $l$-boson pair vacuum state equivalent to
the basis vectors of $O(2l+1)\supset O(3)$
satisfying

$$P_{l}\left\vert
\begin{array}{l}
~[\tau\dot{0}]\\
~(\tau\dot{0})\\
\alpha LM_{L}\\
\end{array}
\right\rangle\equiv
P_{l}\left\vert
\begin{array}{l}
~(\tau\dot{0})\\
\alpha LM_{L}\\
\end{array}
\right\rangle=
0.\eqno(8)$$
 {It follows from this that} once the orthonormal basis vectors
of $U(2l-1)\supset O(2l-1)$ are constructed,
those of $U(2l+1)\supset O(2l+1)\supset O(3)$  {can be}
worked out according to Eq. (7),  {which provides a
recursive procedure for constructing}
basis vectors of $U(2l+1)\supset O(2l+1)\supset O(3)$
from those of $U(2l-1)\supset O(2l-1)$
starting with $l=2$.

\section{Matrix representations of $O(5)$ in the $O_{1}(3)\times U(1)$ basis}
In the following, we use (7) to construct
the basis vectors of $O(5)\supset O_{2}(3)$
from those of the $O(5)\supset O_{1}(3)\otimes U(1)$ as the starting point,
where the quantum numbers of $O_{2}(3)\equiv O(3)$ are exactly those of the angular momentum
of the $d$-boson system, of which the creation operators are expressed as
$\{b^{\dagger}_{-2},b^{\dagger}_{-1},\cdots,b^{\dagger}_{2}\}$.
The procedure involves two steps: (i) Firstly, we construct the basis vectors
of  $O(5)\supset O_{1}(3)\otimes U(1)$ from those
of  $U(5)\supset U(3)\otimes U(2)\supset O_{1}(3)\otimes U(1)$.
(ii) Then, we expand the basis vectors of $O(5)\supset O(3)$
in terms of those of  $O(5)\supset O_{1}(3)\otimes U(1)$.

In this case, generators of $O_{1}(3)$ are written in the canonical
form as

$$l_{+}=\sqrt{2}(b^{\dagger}_{1}b_{0}+b^{\dagger}_{0}b_{-1}),~
l_{-}=(l_{+})^{\dagger},~l_{0}=b^{\dagger}_{1}b_{1}-b^{\dagger}_{-1}b_{-1},\eqno(9)$$
which satisfy the commutation relations

$$[l_{+},~l_{-}]=2l_{0},~[l_{0},~l_{\pm}]=\pm l_{\pm}.\eqno(10)$$
Similarly, generators of $O(3)$ are written as

$$L_{+}=\sqrt{3\over{2}}l_{+}+\sqrt{2}(b^{\dagger}_{2}b_{1}+b^{\dagger}_{-1}b_{-2}),~
L_{-}=(L_{+})^{\dagger},~L_{0}=l_{0}+4J_{0}.\eqno(11)$$

The orthonormal basis vectors of $U(3)\supset O_{1}(3)\supset O_{1}(2)$
and those of the $U(2)\supset U(1)$ are well known~\cite{mos,bie}:

$$\left\vert
\begin{array}{l}
~[r+2\xi~\dot{0}]\cr
~~~~~~r\\
~~~~~m_{r}\\
\end{array}
\right\rangle=\left[
{(2r+1)!!\over{\xi!(2r+2\xi+1)!!}}\right]^{1\over{2}}
P_{1}^{\dagger\xi}
\left\vert
\begin{array}{l}
~r\\
m_{r}\\
\end{array}
\right\rangle=$$
$$\left[
{2^{r+m_{r}}(2r+1)!!(r+m_{r})!(r-m_{r})!r!\over{\xi!(2r+2\xi+1)!!(2r)!}}\right]^{1\over{2}}
P_{1}^{\dagger\xi}
\sum_{x}{b^{\dagger x}_{1}b_{0}^{\dagger r+m_{r}-2x}b_{-1}^{\dagger x-m_{r}}\over{
2^{x}(x-m_{r})!x!(r+m_{r}-2x)!}}\vert 0\rangle
\eqno(12)
$$
for the $U(3)\supset O_{1}(3)\supset O_{1}(2)$, where $\vert 0\rangle$
is the boson vacuum state, and

$$
\left\vert
\begin{array}{l}
~J\\
m_{J}\\
\end{array}
\right\rangle=
{~~~~b^{\dagger J+m_{J}}_{2}b_{-2}^{\dagger J-m_{J}}\over{
\sqrt{(J+m_{J})!(J-m_{J})!}}}\vert 0\rangle\eqno(13)$$
for the $U(2)\supset U(1)$.

According to (7), (12), and (13), the $O(5)\supset O_{1}(3)\otimes U(1)$ basis vectors
may be expanded in terms of those
$U(5)\supset U(3)\otimes U(2)\supset O_{1}(3)\otimes U(1)$
as
$$\left\vert
\begin{array}{l}
(r+2m_{J}+t~0)\\
~~~r~m_{r}, ~m_{J}\\
\end{array}\right\rangle=
\sum_{\xi=0}^{t/2} a^{t,r,m_{J}}_{\xi}
\left[
{(2r+1)!!\over{\xi!(2r+2\xi+1)!!}}\right]^{1\over{2}}
P_{1}^{\dagger\xi}
\left\vert
\begin{array}{l}
[r~\dot{0}]\cr
~~r\\
~~m_{r}\\
\end{array}
\left.
\begin{array}{l}
m_{J}+t/2-\xi\\
~~~~~~m_{J}\\
\end{array}\right.
\right\rangle,\eqno(14)
$$
where $t$ is an even integer, which should satisfy

$$P_{2}\left\vert
\begin{array}{l}
(r+2m_{J}+t~0)\\
~~~r~m_{r}, ~m_{J}\\
\end{array}\right\rangle=
\left(\sqrt{2}b_{2}b_{-2}-P_{1}\right)\left\vert
\begin{array}{l}
(r+t+2m_{J}~0)\\
~~~r~m_{r}, ~m_{J}\\
\end{array}\right\rangle
=0.\eqno(15)$$
Eq. (15) leads to the following relation:

$$a^{t,r,m_{J}}_{\xi+1}=\left[
{(4m_{J}+t-2\xi)(t-2\xi)\over{2(\xi+1)(2r+2\xi+3)}}\right]^{1\over{2}}
a^{t,r,m_{J}}_{\xi}.\eqno(16)$$
Using Eq. (16), we have

$$a^{t,r,m_{J}}_{\xi}=\left[{(4m_{J}+t)!!(2r+1)!!t!!\over{(4m_{J}+t-2\xi)!!(2\xi)!!(2r+2\xi+1)!!(t-2\xi)!!}}\right]^{1\over{2}}
a^{t,r,m_{J}}_{0}.\eqno(17)$$
Substituting (17) into (14), one has

$$
\left\vert
\begin{array}{l}
(r+2m_{J}+t~0)\\
~~~r~m_{r}, ~m_{J}\\
\end{array}\right\rangle=
\sum_{\xi=0}^{t/2}
\left[{(4m_{J}+t)!!(2r+1)!!^2 t!!\over{(4m_{J}+t-2\xi)!!(2\xi)!!(2r+2\xi+1)!!^2 (t-2\xi)!!\xi!
}}\right]^{1\over{2}}\times$$$$
a^{t,r,m_{J}}_{0}P^{\dagger\xi}_{1}
\left\vert
\begin{array}{l}
[r~\dot{0}]\cr
~~r\\
~~m_{r}\\
\end{array}
\left.
\begin{array}{l}
~m_{J}+t/2-\xi\\
~~~~~~m_{J}\\
\end{array}\right.
\right\rangle.\eqno(18)
$$
The normalization condition of (18) leads to the following expression

$$
\left\vert
\begin{array}{l}
~~~(\tau~0)\\
r~m_{r}, ~m_{J}\\
\end{array}\right\rangle=
\sum_{\xi=0}^{t/2}
\left[{(2\tau+1-t)!!(4m_{J}+t)!!(2r+t+1)!! t!!\over{(2\tau+1)!!(4m_{J}+t-2\xi)!!(2\xi)!!(2r+2\xi+1)!! (t-2\xi)!!
}}\right]^{1\over{2}}\times$$$$
\left\vert
\begin{array}{l}
[r+2\xi~0]\\
~~~~r\\
~~~m_{r}\\
\end{array}
\left.
\begin{array}{l}
~m_{J}+t/2-\xi\\
~~~~~~m_{J}\\
\end{array}\right.
\right\rangle=\sum_{\xi=0}^{t/2}b_{\xi}^{\tau,r,m_{J},t}\left\vert
\begin{array}{l}
[r+2\xi~0]\\
~~~~r\\
~~~m_{r}\\
\end{array}
\left.
\begin{array}{l}
~m_{J}+t/2-\xi\\
~~~~~~m_{J}\\
\end{array}\right.
\right\rangle
\eqno(19)
$$
with $\tau=r+2m_{J}+t$.
In derivation of (19), the identity

$$
\sum_{\xi=0}^{t/2}
{(4m_{J}+t)!! t!!\over{(4m_{J}+t-2\xi)!!(2\xi)!!(2r+2\xi+1)!! (t-2\xi)!!
}}
=
{(2\tau+1)!! \over{(2\tau-t+1)!!(2r+t+1)!!
}} \eqno(20)$$ is used, and the overall phase of (19) is thus fixed.
It is clear from the construction of (19) that the branching rule of $O(5)\downarrow O_{1}(3)\otimes U(1)$
for the symmetric irrep $(\tau~0)$ of $O(5)$ is given by

$$r+2m_J=~\tau, ~\tau-2,~\tau-4,~ \cdots, \left\{
\begin{array}{l}
0~ {\rm when}~\tau~ {\rm is~ even},\\
1 ~{\rm when}~ \tau~{\rm is~ odd}.\\
\end{array}\right.\eqno(21)$$

Under the $O(5)\supset O_{1}(3)\times U(1)$ basis, the boson operators
$\{b^{\dagger}_1,b^{\dagger}_0,b^{\dagger}_{-1},b^{\dagger}_{2},b^{\dagger}_{-2}\}$
are rank-$1$ irreducible tensor operators of $O(5)$ with $T^{(10)}_{1\mu;0}=b^{\dagger}_\mu$
for $\mu=1,0,-1$, and $T^{(10)}_{00;\pm{1\over{2}}}=b^{\dagger}_{\pm 2 }$.
Since these irreducible tensor operators appear
in (11), we need matrix elements of them under the
$O(5)\supset O(3)_{1}\times O(2)$ basis in order to make the angular
momentum projection.

By using the explicit expression (19) and Wigner-Eckart theorem, one finds

$$b^{\dagger}_{2}\left\vert
\begin{array}{l}
~~~(\tau~0)\\
r~m_{r}, ~m_{J}\\
\end{array}\right\rangle=\sum_{\xi=0}^{t/2}b_{\xi}^{\tau,r,m_{J},t}
\left\langle
\begin{array}{l}
m_{J}+t/2-\xi\\
~~~~m_{J}\\
\end{array}
\begin{array}{l}
{1/2}\\
{1/2}\\
\end{array}\right\vert\left.
\begin{array}{l}
m_{J}+t/2-\xi+1/2\\
~~~~~m_{J}+1/2\\
\end{array}
\right\rangle\times$$
$$\sqrt{2m_{J}+t-2\xi+1}\left\vert
\begin{array}{l}
[r+2\xi~0]\\
~~~~r\\
~~~m_{r}\\
\end{array}
\begin{array}{l}
~m_{J}+t/2-\xi+1/2\\
~~~~~~m_{J}+1/2\\
\end{array}
\right\rangle=$$$$\sum_{\xi=0}^{t/2}b_{\xi}^{\tau,r,m_{J},t}
\sqrt{{1\over{2}}(4m_{J}+t-2\xi+2)}\left\vert
\begin{array}{l}
[r+2\xi~0]\\
~~~~r\\
~~~m_{r}\\
\end{array}
\begin{array}{l}
~m_{J}+t/2-\xi+1/2\\
~~~~~~m_{J}+1/2\\
\end{array}
\right\rangle,\eqno(22)$$
where $\left\langle
\begin{array}{l}
m_{J}+t/2-\xi\\
~~~~m_{J}\\
\end{array}
\begin{array}{l}
{1/2}\\
{1/2}\\
\end{array}\right\vert\left.
\begin{array}{l}
m_{J}+t/2-\xi+1/2\\
~~~~~m_{J}+1/2\\
\end{array}
\right\rangle$ is the CG coefficient of $U(2)$,
from which we obtain

$$\left\langle
\begin{array}{l}
~~~(\tau+1~0)\\
r~m_{r}, m_{J}+1/2\\
\end{array}
\right\vert b^{\dagger}_{2}\left\vert
\begin{array}{l}
~~~(\tau~0)\\
r~m_{r},~m_{J}\\
\end{array}\right\rangle=\sqrt{(\tau+r+2m_{J}+3)(\tau-r+2m_{J}+2)\over{2(2\tau+3)}}.\eqno(23)$$

While

$$b^{\dagger}_{-2}\left\vert
\begin{array}{l}
~~~(\tau~0)\\
r~m_{r}, ~m_{J}\\
\end{array}\right\rangle=\sum_{\xi=0}^{t/2}b_{\xi}^{\tau,r,m_{J},t}
\sqrt{{1\over{2}}(t-2\xi+2)}\left\vert
\begin{array}{l}
[r+2\xi~0]\\
~~~~r\\
~~~m_{r}\\
\end{array}
\begin{array}{l}
~m_{J}+t/2-\xi+1/2\\
~~~~~~m_{J}-1/2\\
\end{array}
\right\rangle,\eqno(24)$$
from which we have

$$\left\langle
\begin{array}{l}
~~~(\tau+1~0)\\
r~m_{r}, m_{J}-1/2\\
\end{array}
\right\vert b^{\dagger}_{-2}\left\vert
\begin{array}{l}
~~~(\tau~0)\\
r~m_{r},~m_{J}\\
\end{array}\right\rangle=\sqrt{(\tau+r-2m_{J}+3)(\tau-r-2m_{J}+2)\over{2(2\tau+3)}}.\eqno(25)$$

Similarly, we have

$$b^{\dagger}_{1}\left\vert
\begin{array}{l}
~~~(\tau~0)\\
r~m_{r}, ~m_{J}\\
\end{array}\right\rangle=$$$$\sum_{\xi=0}^{t/2}b_{\xi}^{\tau,r,m_{J},t}
\sqrt{(r+m_{r}+1)(r+m_{r}+2)(2r+2\xi+3)\over{2(2r+1)(2r+3)}}\left\vert
\begin{array}{l}
[r+2\xi+1~0]\\
~~~~~r+1\\
~~~~m_{r}+1\\
\end{array}
\begin{array}{l}
~m_{J}+t/2-\xi\\
~~~~~~m_{J}\\
\end{array}
\right\rangle+$$
$$
\sum_{\xi=0}^{t/2}b_{\xi}^{\tau,r,m_{J},t}
\sqrt{(r-m_{r})(r-m_{r}-1)(2\xi+2)\over{2(2r+1)(2r-1)}}\left\vert
\begin{array}{l}
[r+2\xi+1~0]\\
~~~~~r-1\\
~~~~m_{r}+1\\
\end{array}
\begin{array}{l}
~m_{J}+t/2-\xi\\
~~~~~~m_{J}\\
\end{array}
\right\rangle
,\eqno(26)$$
from which we get

$$\left\langle
\begin{array}{l}
~~~(\tau+1~0)\\
r+1~m_{r}+1, m_{J}\\
\end{array}
\right\vert b^{\dagger}_{1}\left\vert
\begin{array}{l}
~~~(\tau~0)\\
r~m_{r},~m_{J}\\
\end{array}\right\rangle=$$$$\sqrt{(\tau+r+2m_{J}+3)(\tau+r-2m_{J}+3)(r+m_{r}+1)(r+m_{r}+2)\over{2(2\tau+3)(2r+3)(2r+1)}}\eqno(27)$$
and
$$\left\langle
\begin{array}{l}
~~~(\tau+1~0)\\
r-1~m_{r}+1, m_{J}\\
\end{array}
\right\vert b^{\dagger}_{1}\left\vert
\begin{array}{l}
~~~(\tau~0)\\
r~m_{r},~m_{J}\\
\end{array}\right\rangle=$$$$\sqrt{(\tau-r+2m_{J}+2)(\tau-r-2m_{J}+2))(r-m_{r})(r-m_{r}-1)\over{2(2\tau+3)(2r+1)(2r-1)}}\eqno(28)$$

By using the similar procedure, we also get

$$\left\langle
\begin{array}{l}
~~~(\tau+1~0)\\
r+1~m_{r}-1, m_{J}\\
\end{array}
\right\vert b^{\dagger}_{-1}\left\vert
\begin{array}{l}
~~~(\tau~0)\\
r~m_{r},~m_{J}\\
\end{array}\right\rangle=$$
$$\sqrt{(\tau+r+2m_{J}+3)(\tau+r-2m_{J}+3)(r-m_{r}+1)(r-m_{r}+2)\over{2(2\tau+3)(2r+3)(2r+1)}},\eqno(29)$$

$$\left\langle
\begin{array}{l}
~~~(\tau+1~0)\\
r-1~m_{r}-1, m_{J}\\
\end{array}
\right\vert b^{\dagger}_{-1}\left\vert
\begin{array}{l}
~~~(\tau~0)\\
r~m_{r},~m_{J}\\
\end{array}\right\rangle=$$
$$\sqrt{(\tau-r+2m_{J}+2)(\tau-r-2m_{J}+2)(r+m_{r})(r+m_{r}-1)\over{2(2\tau+3)(2r+1)(2r-1)}},\eqno(30)$$

$$\left\langle
\begin{array}{l}
~~~(\tau+1~0)\\
r+1~m_{r}, m_{J}\\
\end{array}
\right\vert b^{\dagger}_{0}\left\vert
\begin{array}{l}
~~~(\tau~0)\\
r~m_{r},~m_{J}\\
\end{array}\right\rangle=$$
$$\sqrt{(\tau+r+2m_{J}+3)(\tau+r-2m_{J}+3)(r-m_{r}+1)(r+m_{r}+1)\over{(2\tau+3)(2r+3)(2r+1)}},\eqno(31)$$
and

$$\left\langle
\begin{array}{l}
~~~(\tau+1~0)\\
r-1~m_{r}, m_{J}\\
\end{array}
\right\vert b^{\dagger}_{0}\left\vert
\begin{array}{l}
~~~(\tau~0)\\
r~m_{r},~m_{J}\\
\end{array}\right\rangle=$$
$$-\sqrt{(\tau-r+2m_{J}+2)(\tau-r-2m_{J}+2)(r+m_{r})(r-+m_{r})\over{(2\tau+3)(2r+1)(2r-1)}}.\eqno(32)$$
As is well-known, the matrix elements of single-boson operators (23), (25), and (27)-(32)
are key in deriving matrix elements of the $O(5)$ generators.
Thus, the matrix representations of $O(5)\supset O_{1}(3)\otimes U(1)$  are
completely determined.

\vskip .3cm
Using the Racah factorization lemma~\cite{wybourne},  {which is also called} the generalized Wigner-Eckart theorem, we have

$$\left\langle
\begin{array}{l}
~(\tau+1~0)\\
r^{\prime}~m^{\prime}_{r}, m_{J}\\
\end{array}
\right\vert T^{(10)}_{1\mu;0}\left\vert
\begin{array}{l}
~(\tau~0)\\
r~m_{r},m_{J}\\
\end{array}\right\rangle=\left\langle\begin{array}{c}
(\tau~0)\cr
r,~m_{J}\cr
\end{array}
\begin{array}{c}
(1~0)\cr
1,~0\cr
\end{array}\right\vert\left.
\begin{array}{c}
(\tau+1~0)\cr
r^{\prime},~m_{J}\cr
\end{array}
\right\rangle\left\langle rm_{r},1 \mu\vert r^{\prime}m^{\prime}_{r}\right\rangle
\langle (\tau+1 ~0)\|T^{(10)}\|(\tau ~0)\rangle,\eqno(33)
$$
where $\left\langle\begin{array}{c}
(\tau~0)\cr
r,~m_{J}\cr
\end{array}
\begin{array}{c}
(1~0)\cr
1,~0\cr
\end{array}\right\vert\left.
\begin{array}{c}
(\tau+1~0)\cr
r^{\prime},~m_{J}\cr
\end{array}
\right\rangle$ is the elementary Wigner coefficient or called Isoscalar Factor (ISF) {of} $O(5)\supset O_{1}(3)\otimes U(1)$,
$\left\langle rm_{r},1 \mu\vert r^{\prime}m^{\prime}_{r}\right\rangle$ is the CG coefficient of $O_{1}(3)$,
and $\langle (\tau+1~ 0)\|T^{(10)}\|(\tau~ 0)\rangle$ is the $O(5)$-reduced matrix element
satisfying

$$\langle (\tau^{\prime} 0)\|T^{(10)}\|(\tau 0)\rangle
=\delta_{\tau^{\prime},\tau+1}\sqrt{\tau+1}=
\sqrt{\dim(\tau~0)\over{\dim(\tau+1~0)}}\langle (\tau 0)\|U^{(10)}\|(\tau^{\prime} 0)\rangle,\eqno(34)$$
in which $\dim(\tau~ 0)=(\tau+1)(\tau + 2) (2\tau+3)/6$ is the dimension of the $O(5)$ irrep $(\tau~ 0)$, while
$U^{(10)}_{11,0}=b_{-1}$, $U^{(10)}_{10,0}=-b_{0}$, $U^{(10)}_{1-1,0}=b_{1}$,
$U^{(10)}_{0,1/2}=-b_{-2}$, and $U^{(10)}_{0,-1/2}=-b_{2}$.
Combining Eqs. (33), (34), and the symmetry properties of $O_{1}(3)$ CG coefficients, we have

$$\left\langle\begin{array}{c}
(\tau+1~0)\cr
r^{\prime},~m_{J}\cr
\end{array}
\begin{array}{c}
(1~0)\cr
1,~0\cr
\end{array}\right\vert\left.
\begin{array}{c}
(\tau~0)\cr
r,~m_{J}\cr
\end{array}
\right\rangle=(-1)^{r+1-r^{\prime}}\sqrt{(2r^{\prime}+1)\dim(\tau 0)\over{(2r+1)\dim(\tau+1~0)}}
\left\langle\begin{array}{c}
(\tau~0)\cr
r,~m_{J}\cr
\end{array}
\begin{array}{c}
(1~0)\cr
1,~0\cr
\end{array}\right\vert\left.
\begin{array}{c}
(\tau+1~0)\cr
r^{\prime},~m_{J}\cr
\end{array}
\right\rangle.\eqno(35)
$$
Similarly, we also have

$$\left\langle\begin{array}{c}
(\tau+1~0)\cr
r,~m^{\prime}_{J}\cr
\end{array}
\begin{array}{c}
(1~0)\cr
0,\pm{1/2}\cr
\end{array}\right\vert\left.
\begin{array}{c}
(\tau~0)\cr
r,~m_{J}\cr
\end{array}
\right\rangle=-\sqrt{\dim(\tau 0)\over{\dim(\tau+1~0)}}
\left\langle\begin{array}{c}
(\tau~0)\cr
r,~m_{J}\cr
\end{array}
\begin{array}{c}
(1~0)\cr
0,\mp1/2\cr
\end{array}\right\vert\left.
\begin{array}{c}
(\tau+1~0)\cr
r,~m^{\prime}_{J}\cr
\end{array}
\right\rangle.\eqno(36)
$$

All nonzero elementary Wigner coefficients of the $O(5)\supset O_{1}(3)\otimes U(1)$
are listed in  {Table I. These are useful for calculating matrix elements
of the $O(5)$ irreducible tensor operators
in} the $O(5)\supset O_{1}(3)\otimes U(1)$ basis.

\begin{table}[htb]
\caption[c]{Wigner coefficients
$\left\langle\begin{array}{c}
(\tau^{\prime}~0)\cr
r^{\prime},~m^{\prime}_{J}\cr
\end{array}
\begin{array}{c}
(1~0)\cr
\mu,~m\cr
\end{array}\right\vert\left.
\begin{array}{c}
(\tau~0)\cr
r,~m_{J}\cr
\end{array}
\right\rangle$
{of} $O(5)\supset O_{1}(3)\otimes U(1)$.}
\begin{center}
\begin{tabular}{cccccc}
\hline
$\tau^{\prime}$$\left\backslash
\begin{array}{c}
r^{\prime},~m^{\prime}_J\cr
\mu,~m\cr
\end{array}\right.$
&{$\begin{array}{c}r,~m_J-1/2\cr
 0,~1/2\cr
 \end{array}$}
 &{$\begin{array}{c}r,~m_J+1/2\cr
 0,~-1/2\cr
 \end{array}$}
 \\
\hline
\\
$\tau+1$~~~~~~~~~~~~~~~~~&$-\left[{(\tau+r-2m_J+3)(\tau-r-2m_J+2)\over{2(\tau+3)(2\tau+5)}}\right]^{1\over{2}}$
&$-\left[{(\tau+r+2m_J+3)(\tau-r+2m_J+2)\over{2(\tau+3)(2\tau+5)}}\right]^{1\over{2}}$\\
\\
$\tau-1$~~~~~~~~~~~~~~~~~ &$\left[{(\tau+r+2m_J+1)(\tau-r+2m_J)\over{2\tau(2\tau+1)}}\right]^{1\over{2}}$
&$\left[{(\tau+r-2m_J+1)(\tau-r-2m_J)\over{2\tau(2\tau+1)}}\right]^{1\over{2}}$\\
\\
\hline
\hline
$\tau^{\prime}$$\left\backslash
\begin{array}{c}
r^{\prime},~m^{\prime}_J\cr
\mu,~m\cr
\end{array}\right.$
&{$\begin{array}{c}r-1,~m_J\cr
 1,~0\cr
 \end{array}$}
 &{$\begin{array}{c}r+1,~m_J\cr
 1,~0\cr
 \end{array}$}
 \\
\hline
\\
$\tau+1$~~~~~~~~~~~~~~~~~&$\left[{(\tau-r+2m_J+2)(\tau-r-2m_J+2)r\over{(\tau+3)(2\tau+5)(2r+1)}}\right]^{1\over{2}}$
&$\left[{(\tau+r+2m_J+3)(\tau+r-2m_J+3)(r+1)\over{(\tau+3)(2\tau+5)(2r+1)}}\right]^{1\over{2}}$\\
\\
$\tau-1$~~~~~~~~~~~~~~~~~ &$\left[{(\tau+r+2m_J+1)(\tau+r-2m_J+1)r\over{\tau(2\tau+1)(2r+1)}}\right]^{1\over{2}}$
&$\left[{(\tau-r+2m_J)(\tau-r-2m_J)(r+1)\over{\tau(2\tau+1)(2r+1)}}\right]^{1\over{2}}$\\
\\
\hline\hline
\end{tabular}
\end{center}
\end{table}

\section{The basis vectors of $O(5)\supset O(3)$}

The basis vectors of $O(5)\supset O(3)\supset O(2)$ can now be expanded in terms of
those of the  $O(5)\supset O_{1}(3)\otimes U(1)$
with the
restriction $m_{r} + 4m_{J}=M_{L}$. For a given angular momentum quantum number $L=2\tau-k$ with $M_{L}=L$,
the quantum numbers of the $O_{1}(3)\supset O_{1}(2)$ and that of $U(1)$
may be parameterized as

$$\left\vert
\begin{array}{l}
~~~~~~~~~~~(\tau~ 0)\\
\zeta, L= M_{L}=2\tau-k\\
\end{array}\right\rangle=
\sum_{q,t} c^{(\zeta)}_{q,t}(\tau,k)
\left\vert
\begin{array}{l}
~~~~~~~~~~~~~~~~~~~~~(\tau~0)\\
k-q,k-2q+2t;~
(\tau-k+q-t)/2\\
\end{array}
\right\rangle,\eqno(37)
$$
where $\zeta$ is the multiplicity label needed in the reduction  $O(5)\downarrow O(3)$,
which will be omitted if the reduction is simple,
$c^{(\zeta)}_{q,t}(\tau,k)$ is the corresponding expansion coefficient, and $k=0,1,2,\cdots,2\tau$.
(37) should satisfy

$$L_{+}\left\vert
\begin{array}{l}
~~~~~~~~~~~(\tau~0)\\
\zeta, L=M_{L}=2\tau-k\\
\end{array}\right\rangle=\left(\sqrt{3\over{2}}l_{+}+\sqrt{2}(b^{\dagger}_{2}b_{1}+b^{\dagger}_{-1}b_{-2})
\right)\left\vert
\begin{array}{l}
~~~~~~~~~~~(\tau~0)\\
\zeta, L=M_{L}=2\tau-k\\
\end{array}\right\rangle=0.\eqno(38)$$

According to the Racah factorization lemma~\cite{wybourne}, by using the $O(5)$-reduced matrix element (34) and the
Wigner coefficients shown in Table I,
it can easily be proven that the condition (38)
leads to the following three-term recurrence relation for the expansion coefficients $c^{(\zeta)}_{q,t}(\tau,k)$ needed in (37):

$$\left[{(2k-3q+2t+2)(2k-3q+2t+3)(2k-2q+t+3)(2\tau-2k+2q-t)\over{(2k-2q+1)(2k-2q+3)}}\right]^{1\over{2}}c^{(\zeta)}_{q-1,t}(\tau,k)
+$$$$\left[3(q-2t)(2k-3q+2t+1)\right]^{1\over{2}}c^{(\zeta)}_{q,t}(\tau,k)+
\left[{(q-2t-1)(q-2t-2)(t+2)(2\tau-t+1)\over{(2k-2q-1)(2k-2q+1)}}\right]^{1\over{2}}c^{(\zeta)}_{q+1,t+2}(\tau,k)=0.\eqno(39)$$

The boundary conditions for integers  $q$ and even integer $t$ can be obtained
from the allowed quantum number $m_{r}=k-2q+2t$ of $O_{1}(2)$ under the
reduction of $O_{1}(3)\supset O_{1}(2)$ and allowed $m_{J}=(n-k+q-t)/2$ of $U(1)$
under the reduction of $U(2)\supset U(1)$ according to (37), which can be specified
as

$$k - q \geq\vert k - 2 q + 2 t\vert\eqno(40)$$
with $0\leq q\leq k$ and $0\leq t\leq {\rm Int}[k/2]$, where
${\rm Int}[x]$ is the integer part of $x$. A set of allowed ($q$, $t$)
combinations for given $k$ are listed in Table II for $0\leq k\leq 10$,
which is generated by a simple Mathematica code according to (40).

\begin{table}[htb]
\caption{Allowed ($q$, $t$) combinations in the basis vectors (37)
of $O(5)\supset O(3)$ for $L=2\tau-k$ expanded in terms of those of $O_{1}(3)\otimes U(1)$
and the corresponding multiplicity ${\rm Multi}(\tau,k)$ for $\tau\geq 10$ and $k\leq 10$,
where $d$ is the total number of terms needed in the expansion.}
\begin{tabular}{cccccc}
\hline
$k$~&$(q,t)$&$d$ &${\rm Multi}(\tau,k)$\\
\hline
$0$~~&$(0,0)$ &$1$&$1$\\
$1$~~&- &$0$&$0$\\
$2$~~&$(0,0),(1,0)$&$2$&$1$\\
$3$~~&$(0,0),(1,0),(2,0)$&$3$&$1$\\
$4$~~&$(0,0),(1,0),(2,0),(4,2)$&$4$&$1$\\
$5$~~&$(0,0),(1,0),(2,0),(3,0),(4,2)$&$5$&$1$\\
$6$~~&$(0,0),(1,0),(2,0),(3,0),(4,0),(4,2),(5,2)$&$7$&$2$\\
$7$~~&$(0,0),(1,0),(2,0),(3,0),(4,0),(4,2),(5,2),(6,2)$&$8$&$1$\\
$8$~~&$(0,0),(1,0),(2,0),(3,0),(4,0),(4,2),(5,0),(5,2),(6,2),(8,4)$&$10$&$2$\\
$9$~~&$(0,0),(1,0),(2,0),(3,0),(4,0),(4,2),(5,0),(5,2),(6,0),(6,2),(7,2),(8,4)$&$12$&$2$\\
$10$~~&$(0,0),(1,0),(2,0),(3,0),(4,0),(4,2),(5,0),(5,2),(6,0),(6,2),(7,2),(8,2),(8,4),(9,4)$&$14$&$2$\\
\hline\hline
\end{tabular}
\end{table}

Practically, one can construct a matrix equation of (39) with

$${\bf P}(\tau,k){\bf c}^{(\zeta)}(\tau,k)=\lambda {\bf c}^{(\zeta)}(\tau,k),\eqno(41)$$
where the transpose of ${\bf c}^{(\zeta)}(\tau,k)$ is arranged as $\left({{\bf c}^{(\zeta)}(\tau,k)}\right)^{\rm T}=(c^{(\zeta)}_{0,0}(\tau,k),
c^{(\zeta)}_{1,0}(\tau,k),\cdots, c^{(\zeta)}_{4,0}(\tau,k), c^{(\zeta)}_{4,2}(\tau,k),\cdots)$, of which some examples are shown in Table II.
Entries of the angular momentum projection matrix ${\bf P}(\tau,k)$ can easily be read out from Eq. (39).
The components of eigenvector ${\bf c}^{(\zeta)}(\tau,k)$ corresponding to $\lambda=0$
provide the expansion coefficients $\{c^{(\zeta)}_{q,t}(\tau,k)\}$ of (37).
Once the matrix ${\bf P}(\tau,k)$  is constructed, it can be verified that the
number of $\lambda=0$ solutions of Eq. (41) for sufficiently large $\tau$ equals exactly to the number of
rows of ${\bf P}(\tau,k)$ with all entries zero. However, some entries of ${\bf P}(\tau,k)$
will be zero or become complex for some specific values of $\tau$. In such cases,  nonzero solution of $\{c^{(\zeta)}_{q,t}(\tau,k)\}$
does not exist, which will be examined for $\tau\leq 8$ cases separately in the following.
Furthermore, $\left({{\bf c}^{(\zeta^{\prime})}(\tau,k)}\right)^{\rm T}\cdot{\bf c}^{(\zeta)}(\tau,k)\neq0$
when the multiplicity is greater than $1$ mainly because the projection matrix
${\bf P}(\tau,k)$ is nonsymmetric. Therefore, the basis vectors (37) constructed from the expansion coefficients
obtained according to (39) are non-orthogonal with respect to the multiplicity label $\zeta$.
The Gram-Schmidt process may be adopted in order to construct orthonormalized
basis vectors of $O(5)\supset O(3)$.
\vskip .3cm
On the other hand, for given $L=2\tau-k$ of $O(3)$, the number of $\lambda=0$ solutions, ${\rm Multi}(\tau,k)$, of Eq. (41)
with $\zeta=1,2,\cdots, {\rm Multi}(\tau,k)$ equals exactly to the
multiplicity in the reduction  $O(5)\downarrow O(3)$
for the symmetric irrep $(\tau~0)$ of $O(5)$, which
may be calculated in the following way:
Let $Q_{\tau}(k)$ be the number of different $\tau$-partitions of
the positive integer $k$ with $k=\sum_{i=1}^{\tau}\xi_{i}$, where
$4\geq \xi_{1}\geq\xi_{2}\geq\cdots\geq\xi_{\tau}\geq 0$.
Then, ${\rm Multi}(\tau,k)
=Q_{\tau}(k)+Q_{\tau-2}(k-5)-Q_{\tau}(k-1)-Q_{\tau-2}(k-4)$,
where $Q_{\tau}(0)=1$ and $Q_{\tau}(v)=0$ when $v<0$
may be defined for convenience in the computation.
The corresponding  ${\rm Multi}(\tau,k)$ for given $k$ and  $\tau\geq 10$
are also shown in the last column of Table II.

\vskip .3cm
In the following, we list some ${\bf P}(\tau,k)$ matrices and the corresponding
expansion coefficients $\{c^{(\zeta)}_{q,t}(\tau,k)\}$. There is always a freedom
in choosing the global phase. In our calculation, we always set $c_{0,0}(\tau,k)>0$,
while the relative phase is completely determined by the eigen-equation (41).
\vskip .3cm

When $k=0$, ${\bf P}(\tau,0)=0$ with $c_{0,0}(\tau,0)=1$, which is trivial corresponding
to one unique highest weight state of the symmetric irrep $(\tau~0)$  of
$O(5)\supset O(3)$ with $L=2\tau$.
When $k=1$,  ${\bf P}(\tau,1)=2\sqrt{3}$ which requires $c_{0,0}(\tau,1)=0$.
Namely, there is no basis vector for the symmetric irrep $(\tau~0)$ of
$O(5)\supset O(3)$ with $L=2\tau-1$. When $k=2$,
$${\bf P}(\tau,2)=\left(\begin{matrix}
0 &0\cr
2\sqrt{2\tau-2} &\sqrt{6}\cr
\end{matrix}\right)\eqno(42)$$
with $\left({\bf c}(\tau,2)\right)^{\rm T}=\left( c_{0,0}(\tau,2),c_{1,0}(\tau,2)\right)$.
Since there is one row with all entries zero, the multiplicity
of $L=2\tau-2$ is ${\rm Multi}(\tau,2)=1$ for $\tau>1$.
The normalized expansion coefficients are
$c_{0,0}(\tau,2)=\sqrt{3\over{4\tau-1}}$, $c_{1,0}(\tau,2)=-\sqrt{4(\tau-1)\over{4\tau-1}}$
for $\tau>1$. Though arbitrary $c_{0,0}(\tau,2)$ is  {a} possible
solution of (42) when $\tau=1$, only ${\bf c}(\tau,2)=0$ is valid
according to the branching rule of $O(5)\downarrow O(3)$.

For $k=3$,
$${\bf P}(\tau,3)=\left(\begin{matrix}
0 &0 &0\cr
\sqrt{6(2\tau-4)} &2\sqrt{3} &0\cr
0 &\sqrt{2(2\tau-2)} &\sqrt{6}\cr
\end{matrix}\right).\eqno(43)$$
Since there is one row with all entries zero in (43) when  $\tau>2$, the multiplicity
of $L=2\tau-3$ is ${\rm Multi}(\tau,3)=1$ for $\tau>2$.
The normalized nonzero expansion coefficients corresponding
to $\lambda=0$ are
$c_{0,0}(\tau,3)=\sqrt{3\over{(2\tau-1)(\tau-1)}}$, $c_{1,0}(\tau,3)=-\sqrt{3(\tau-2)\over{(2\tau-1)(\tau-1)}}$,
$c_{2,0}(\tau,3)=\sqrt{2(\tau-2)\over{2\tau-1}}$
for $\tau>2$. Though arbitrary $c_{0,0}(\tau,3)$ is  {a} possible
solution of (43) when $\tau=2$, only ${\bf c}(\tau,3)=0$ is valid
according to the branching rule of $O(5)\downarrow O(3)$.

For $k=4$,
$${\bf P}(\tau,4)=\left(\begin{matrix}
0 &0 &0&0\cr
\sqrt{8(2\tau-6)} &2\sqrt{3} &0 &0\cr
0 &\sqrt{2(2\tau-4)} &3\sqrt{2} &0\cr
0&0 &\sqrt{4(\tau-1)\over{3}} &\sqrt{4(2\tau+1)\over{3}}\cr
\end{matrix}\right).\eqno(44)$$
Since there is one row with all entries zero in (44)  when  $\tau>3$, the multiplicity
of $L=2\tau-4$ is ${\rm Multi}(\tau,4)=1$ for $\tau>3$.
The normalized nonzero expansion coefficients corresponding
to $\lambda=0$ are
$c_{0,0}(\tau,4)=\sqrt{27(2\tau+1)\over{(2\tau-3)(4\tau-3)(4\tau-5)}}$,
$c_{1,0}(\tau,4)=-\sqrt{24(2\tau+1)(\tau-3)\over{(2\tau-3)(4\tau-3)(4\tau-5)}}$,
$c_{2,0}(\tau,4)=\sqrt{32(2\tau+1)(\tau-2)(\tau-3)\over{3(2\tau-3)(4\tau-3)(4\tau-5)}}$,
$c_{4,2}(\tau,4)=-\sqrt{32(\tau-1)(\tau-2)(\tau-3)\over{3(2\tau-3)(4\tau-3)(4\tau-5)}}$
for $\tau> 3$. Similar to the $k=3$ case, though arbitrary $c_{0,0}(\tau)$ is  {a} possible
solution of (44) when $\tau=3$, only ${\bf c}(\tau,4)=0$ is valid
according to the branching rule of $O(5)\downarrow O(3)$.
Namely, $L=2$ does not occur in the reduction $(3~0)\downarrow L$.
Moreover, one entry in (44) becomes complex when $\tau=2$
which must be ruled out since complex solutions obviously violate the
branching rule of $O(5)\downarrow O(3)$.

For $k=5$,
$${\bf P}(\tau,5)=\left(\begin{matrix}
0 &0 &0&0&0\cr
\sqrt{10(2\tau-8)} &2\sqrt{6} &0 &0\cr
0 &\sqrt{6(2\tau-6)} &\sqrt{30} &0&0\cr
0&0 &\sqrt{12(2\tau-4)\over{5}} &3\sqrt{2}&\sqrt{4(2\tau+1)\over{15}}\cr
0&0&0&0&\sqrt{20(\tau-1)\over{3}}\cr
\end{matrix}\right).\eqno(45)$$
Since there is also one row with the entries all zero in (45) when $\tau>4$, the multiplicity
of $L=2\tau-5$ is ${\rm Multi}(\tau,5)=1$ for $\tau>4$.
The normalized nonzero expansion coefficients corresponding
to $\lambda=0$ are
$c_{0,0}(\tau,5)=\sqrt{90\over{(2\tau-3)(4\tau-7)(\tau-2)}}$,
$c_{1,0}(\tau,5)=-\sqrt{75(\tau-4)\over{(2\tau-3)(4\tau-7)(\tau-2)}}$,
$c_{2,0}(\tau,5)=\sqrt{30(\tau-3)(\tau-4)\over{(2\tau-3)(4\tau-7)(\tau-2)}}$,
$c_{3,0}(\tau,5)=-\sqrt{8(\tau-3)(\tau-4)\over{(2\tau-3)(4\tau-7)}}$,
$c_{4,2}(\tau,5)=0$
for $\tau> 4$. Similar to the discussions shown in the previous
examples, only $\tau> 4$ is allowed for the $k=5$ case.

For $k=6$,
$${\bf P}(\tau,6)=\left(\begin{matrix}
0 &0 &0&0&0&0&0\cr
\sqrt{12(2\tau-10)} &\sqrt{30} &0 &0&0&0\cr
0  &\sqrt{8(2\tau-8)} &\sqrt{42}&0 &0 &0&0\cr
0 &0 &\sqrt{30(2\tau-6)\over{7}} &6 &0 &\sqrt{4(2\tau+1)\over{35}}&0 \cr
0&0 &0 &\sqrt{72(2\tau-4)\over{5}}&2\sqrt{3} &0 &\sqrt{4(2\tau+1)\over{5}}\cr
0  &0 &0 &0 &0 &0&0\cr
0&0&0&0&0 &\sqrt{28(2\tau-4)\over{5}} &\sqrt{6}\cr
\end{matrix}\right).\eqno(46)$$
Since there are two rows with all entries zero in (44) when $\tau>5$, the multiplicity
of $L=2\tau-6$ is ${\rm Multi}(\tau,6)=2$ for $\tau>5$.
The normalized nonzero expansion coefficients corresponding
to $\lambda=0$ in this case are

\begin{center}
$c^{(\zeta=1)}_{0,0}(\tau,6)=\sqrt{2205\over{f(\tau)}},~
c^{(\zeta=1)}_{1,0}(\tau,6)=-\sqrt{1764(\tau-5)\over{f(\tau)}},~
c^{(\zeta=1)}_{2,0}(\tau,6)=\sqrt{672(\tau-4)(\tau-5)\over{f(\tau)}}$,\\
\vskip .3cm
$c^{(\zeta=1)}_{3,0}(\tau,6)=-\sqrt{160(\tau-3)(\tau-4)(\tau-5)\over{f(\tau)}}$,~
$c^{(\zeta=1)}_{4,0}(\tau,6)=\sqrt{32(\tau-2)(\tau-3)(\tau-4)(\tau-5)\over{f(\tau)}},~
c^{(\zeta=1)}_{4,2}(\tau,6)=0,~~c^{(\zeta=1)}_{5,2}(\tau)=0$,\\
\end{center}
where $f(\tau)=32\tau^4-288\tau^3+1024\tau^2-1692\tau+1065$,
and
\begin{center}
$c^{(\zeta=2)}_{0,0}(\tau,6)=\sqrt{405(2\tau+1)\over{128\tau^4-1376\tau^3+5608\tau^2-10042\tau+6465}},~~
c^{(\zeta=2)}_{1,0}(\tau,6)=-\sqrt{324(2\tau+1)(\tau-5)\over{128\tau^4-1376\tau^3+5608\tau^2-10042\tau+6465}}$,\\
\vskip .2cm
$c^{(\zeta=2)}_{2,0}(\tau,6)=\sqrt{864(2\tau+1)(\tau-4)(\tau-5)\over{7(128\tau^4-1376\tau^3+5608\tau^2-10042\tau+6465)}},~
c^{(\zeta=2)}_{3,0}(\tau,6)=-\sqrt{128(2\tau+1)(\tau-3)(\tau-4)(\tau-5)\over{5(128\tau^4-1376\tau^3+5608\tau^2-10042\tau+6465)}}$,\\
\vskip .2cm
$c^{(\zeta=2)}_{4,0}(\tau,6)=0$,~
$c^{(\zeta=2)}_{4,2}(\tau,6)=-\sqrt{288(\tau-3)(\tau-4)(\tau-5)\over{7(128\tau^4-1376\tau^3+5608\tau^2-10042\tau+6465)}}$,\\
\vskip .2cm
$c^{(\zeta=2)}_{5,2}(\tau,6)=\sqrt{384(\tau-2)(\tau-3)(\tau-4)(\tau-5)\over{5(128\tau^4-1376\tau^3+5608\tau^2-10042\tau+6465)}}$.
\end{center}
One can verify the $\left({\bf c}^{(\zeta=1)}(\tau,6)\right)^{T}\cdot {\bf c}^{(\zeta=2)}(\tau,6)\neq0$.
After the Gram-Schmidt orthonormalization, we have

\begin{center}

${\bar{{\bf c}}}^{\chi=1}(\tau,6)={\bf c}^{(\zeta=1)}(\tau,6)$;\\
\vskip .2cm
$\bar{c}^{\chi=2}_{0,0}(\tau,6)=-{12(3\tau-5)\sqrt{10(2\tau+1)(\tau-3)(\tau-4)(\tau-5)}\over{
\sqrt{(4\tau-5)(4\tau-7)(4\tau-9)(2\tau-5)(\tau-2)f(\tau)}}}$,\\
\vskip .2cm
$\bar{c}^{\chi=2}_{1,0}(\tau,6)={24(\tau-5)(3\tau-5)\sqrt{2(2\tau+1)(\tau-3)(\tau-4)}\over{
\sqrt{(4\tau-5)(4\tau-7)(4\tau-9)(2\tau-5)(\tau-2)f(\tau)}}}$,\\
\vskip .2cm
$\bar{c}^{\chi=2}_{2,0}(\tau,6)=-{32(3\tau-5)(\tau-4)(\tau-5)\sqrt{3(2\tau+1)(\tau-3)}\over{
\sqrt{7(4\tau-5)(4\tau-7)(4\tau-9)(2\tau-5)(\tau-2)f(\tau)}}}$,\\
\vskip .2cm
$\bar{c}^{\chi=2}_{3,0}(\tau,6)={(64\tau^4-896\tau^3+4448\tau^2-9244\tau+6705)\sqrt{(2\tau+1)}\over{
\sqrt{5(4\tau-5)(4\tau-7)(4\tau-9)(2\tau-5)(\tau-2)f(\tau)}}}$,\\
\vskip .2cm
$\bar{c}^{\chi=2}_{4,0}(\tau,6)={\sqrt{(2\tau+1)}(64\tau^3-480\tau^2+1172\tau-915)\over{
\sqrt{(4\tau-5)(4\tau-7)(4\tau-9)(2\tau-5)f(\tau)}}}$,\\
\vskip .2cm
$\bar{c}^{\chi=2}_{4,2}(\tau,6)={3\sqrt{f(\tau)}\over{
\sqrt{7(4\tau-5)(4\tau-7)(4\tau-9)(2\tau-5)(\tau-2)}}}$,\\
\vskip .2cm
$\bar{c}^{\chi=2}_{5,2}(\tau,6)=-{2\sqrt{3f(\tau)}\over{
\sqrt{5(4\tau-5)(4\tau-7)(4\tau-9)(2\tau-5)}}}$.\\
\end{center}

Instead of ${\bf c}^{(\zeta)}(\tau,6)$, the basis vectors (37) with the expansion coefficients
$\bar{\bf c}^{(\chi)}(\tau,6)$ for $\tau>5$ are orthonormal with respect to the new multiplicity label $\chi$.

Similar to discussions in previous examples,
the expansion coefficients $c_{0,0}(\tau,6)$, $c_{1,0}(\tau,6)$, and $c_{2,0}(\tau,6)$ become zero
when $3\leq\tau\leq 5$ for $L=2\tau-6$. The effective projection matrix ${\bf P}(\tau,6)$
in this case becomes

$${\bf P}(\tau,6)=\left(\begin{matrix}
6 &0 &\sqrt{4(2\tau+1)\over{35}}&0 \cr
\sqrt{72(2\tau-4)\over{5}}&2\sqrt{3} &0 &\sqrt{4(2\tau+1)\over{5}}\cr
0 &0 &0&0\cr
0 &0&\sqrt{28(2\tau-4)\over{5}} &\sqrt{6}\cr
\end{matrix}\right)\eqno(47)$$
with the remaining nonzero components of ${\bf c}^{T}(\tau,6)$ arranged as
$\{ c_{3,0}(\tau,6),c_{4,0}(\tau,6),c_{4,2}(\tau,6), c_{5,2}(\tau,6)\}$.
Obviously, the multiplicity of $L=2\tau-6$ when $3\leq\tau\leq 5$ becomes $1$
with the normalized nonzero expansion coefficients\\

\begin{center}
$c_{3,0}(\tau,6)=\sqrt{2\tau+1\over{5(18\tau^2+91\tau-190)}}$,~
$c_{4,0}(\tau,6)=-\sqrt{9(2\tau+1)(\tau-2)\over{18\tau^2+91\tau-190}}$,\\
$c_{4,2}(\tau,6)=-{3\sqrt{7}\over{\sqrt{18\tau^2+91\tau-190}}}$,~
$c_{5,2}(\tau,6)={14\sqrt{3(\tau-2)}\over{\sqrt{5(18\tau^2+91\tau-190)}}}.$
\end{center}

\vskip .3cm
For $k=7$
$$\tiny{\bf P}(\tau,7)=\left(\begin{matrix}
0 &0 &0&0&0&0&0&0\cr
\sqrt{14(2\tau-12)} &6 &0 &0 &0&0&0\cr
0  &\sqrt{10(2\tau-10)} &3\sqrt{6}&0 &0 &0 &0&0\cr
0  &0 &{2\over{3}}\sqrt{14(2\tau-8)} &3\sqrt{6}&0 &{2\over{3}}\sqrt{2\tau+1\over{7}}  &0&0\cr
0 &0 &0 &2\sqrt{5(2\tau-6)\over{7}} &6 &0 &\sqrt{12(2\tau+1)\over{35}}&0 \cr
0&0 &0 &0&\sqrt{4(2\tau-4)\over{5}}&0 &0 &\sqrt{8(2\tau+1)\over{5}}\cr
0  &0 &0 &0 &0 &3\sqrt{6(2\tau-6)\over{7}} &2\sqrt{3} &0\cr
0&0&0&0&0 &0 &\sqrt{14(2\tau-4)\over{5}} &\sqrt{6}\cr
\end{matrix}\right).\eqno(48)$$
Since there is also one row with all entries zero in (48) when $\tau>6$, the multiplicity
of $L=2\tau-7$ is ${\rm Multi}(\tau,7)=1$ for $\tau>6$.
The normalized nonzero expansion coefficients corresponding
to $\lambda=0$ are

\begin{center}
$c_{0,0}(\tau,7)=9\sqrt{35(2\tau+1)\over{(4\tau-11)(4\tau-9)(2\tau-3)(2\tau-5)(\tau-3)}}$,~
$c_{1,0}(\tau,7)=-21\sqrt{5(2\tau+1)(\tau-6)\over{(4\tau-11)(4\tau-9)(2\tau-3)(2\tau-5)(\tau-3)}}$,\\
\vskip .2cm
$c_{2,0}(\tau,7)=35\sqrt{2(2\tau+1)(\tau-5)(\tau-6)\over{3(4\tau-11)(4\tau-9)(2\tau-3)(2\tau-5)(\tau-3)}}$,
$c_{3,0}(\tau,7)=-36\sqrt{(2\tau+1)(\tau-4)(\tau-5)(\tau-6)\over{7(4\tau-11)(4\tau-9)(2\tau-3)(2\tau-5)(\tau-3)}}$,\\
\vskip .2cm
$c_{4,0}(\tau,7)=8\sqrt{2(2\tau+1)(\tau-4)(\tau-5)(\tau-6)\over{5(4\tau-11)(4\tau-9)(2\tau-3)(2\tau-5)}}$,
$c_{4,2}(\tau,7)=-4\sqrt{2(\tau-4)(\tau-5)(\tau-6)\over{3(4\tau-11)(4\tau-9)(2\tau-3)(2\tau-5)(\tau-3)}}$,\\
\vskip .2cm
$c_{5,2}(\tau,7)=4\sqrt{6(\tau-4)(\tau-5)(\tau-6)\over{7(4\tau-11)(4\tau-9)(2\tau-3)(2\tau-5)}}$,
$c_{6,2}(\tau,7)=-8\sqrt{(\tau-2)(\tau-4)(\tau-5)(\tau-6)\over{5(4\tau-11)(4\tau-9)(2\tau-3)(2\tau-5)}}$.
\end{center}
Using the similar procedure exemplified in the previous $k=6$ case, one can verify
that no nonzero solution exists for $L=2\tau-7$ with $4\leq\tau\leq 6$.

\vskip .3cm
For $k=8$,
$$\tiny{\bf P}(\tau,8)=\left(\begin{matrix}
0 &0 &0 &0&0&0&0&0&0&0\cr
{4\sqrt{2\tau-14}} &\sqrt{42} &0&0 &0 &0 &0&0&0&0\cr
0 &\sqrt{12(2\tau-12)} &\sqrt{66} &0 &0 &0 &0&0&0&0\cr
0 &0 &\sqrt{90(2\tau-10)\over{11}} &6\sqrt{2}&0 &\sqrt{4(2\tau+1)\over{99}} &0 &0 &0&0\cr
0  &0 &0 &\sqrt{14(2\tau-8)\over{3}} &\sqrt{60}&0 &0 &\sqrt{4(2\tau+1)\over{21}}  &0&0\cr
0  &0 &0 &0 &0 &0 &0 &0 &0 &0\cr
0 &0 &0 &0 &\sqrt{12(2\tau-6)\over{7}} &0 &\sqrt{30} &0 &\sqrt{24(2\tau+1)\over{35}}&0 \cr
0&0 &0 &0&0 &\sqrt{88(2\tau-8)\over{9}}&0 &3\sqrt{2}&0 &0\cr
0  &0 &0 &0 &0 &0 &0 &\sqrt{36(2\tau-6)\over{7}} &3\sqrt{2} &0\cr
0&0&0&0&0 &0 &0&0 &\sqrt{14(2\tau-4)\over{15}} &\sqrt{8(2\tau-1)\over{3}}\cr
\end{matrix}\right).\eqno(49)$$
Since there is two rows with all entries zero in (49) when $\tau>7$, the multiplicity
of $L=2\tau-8$ is ${\rm Multi}(\tau,8)=2$ for $\tau>7$.
The normalized nonzero expansion coefficients corresponding
to $\lambda=0$ are

\begin{center}
$c^{(\zeta=1)}_{0,0}(\tau,8)=-\sqrt{114345\over{(4\tau-13)(32\tau^4-416\tau^3+2128\tau^2-5044\tau+4515)}}$,~
$c^{(\zeta=1)}_{1,0}(\tau,8)=\sqrt{87120(\tau-7)\over{(4\tau-13)(32\tau^4-416\tau^3+2128\tau^2-5044\tau+4515)}}$,\\
\vskip .2cm
$c^{(\zeta=1)}_{2,0}(\tau,8)=-\sqrt{31680(\tau-6)(\tau-7)\over{(4\tau-13)(32\tau^4-416\tau^3+2128\tau^2-5044\tau+4515)}}$,~
$c^{(\zeta=1)}_{3,0}(\tau,8)=\sqrt{7200(\tau-5)(\tau-6)(\tau-7)\over{(4\tau-13)(32\tau^4-416\tau^3+2128\tau^2-5044\tau+4515)}}$,\\
\vskip .2cm
$c^{(\zeta=1)}_{4,0}(\tau,8)=-\sqrt{1120(\tau-4)(\tau-5)(\tau-6)(\tau-7)\over{(4\tau-13)(32\tau^4-416\tau^3+2128\tau^2-5044\tau+4515)}}$,~
$c^{(\zeta=1)}_{4,2}(\tau,8)=0$,\\
\vskip .2cm
$c^{(\zeta=1)}_{5,0}(\tau,8)=\sqrt{64(\tau-3)(\tau-4)(\tau-5)(\tau-6)(\tau-7)\over{(4\tau-13)(32\tau^4-416\tau^3+2128\tau^2-5044\tau+4515)}}$,~
$c^{(\zeta=1)}_{5,2}(\tau,8)=0$,~
$c^{(\zeta=1)}_{6,2}(\tau,8)=0$,~
$c^{(\zeta=1)}_{8,4}(\tau,8)=0$;
\end{center}
and
\begin{center}
$c^{(\zeta=2)}_{0,0}(\tau,8)=\sqrt{8505(2\tau+1)(2\tau-1)\over{(4\tau-13)(256\tau^5-4800\tau^4+35248\tau^3-122724\tau^2+195220\tau-110355)}}$,\\
\vskip .2cm
$c^{(\zeta=2)}_{1,0}(\tau,8)=-\sqrt{6480(2\tau+1)(2\tau-1)(\tau-7)\over{(4\tau-13)(256\tau^5-4800\tau^4+35248\tau^3-122724\tau^2+195220\tau-110355)}}$,\\
\vskip .2cm
$c^{(\zeta=2)}_{2,0}(\tau,8)=\sqrt{25920(2\tau+1)(2\tau-1)(\tau-6)(\tau-7)\over{11(4\tau-13)(256\tau^5-4800\tau^4+35248\tau^3-122724\tau^2+195220\tau-110355)}}$,\\
\vskip .2cm
$c^{(\zeta=2)}_{3,0}(\tau,8)=-\sqrt{512(2\tau+1)(2\tau-1)(\tau-5)(\tau-6)(\tau-7)\over{(4\tau-13)(256\tau^5-4800\tau^4+35248\tau^3-122724\tau^2+195220\tau-110355)}}$,\\
\vskip .2cm
$c^{(\zeta=2)}_{4,0}(\tau,8)=\sqrt{2048(2\tau+1)(2\tau-1)(\tau-4)(\tau-5)(\tau-6)(\tau-7)\over{35(4\tau-13)(256\tau^5-4800\tau^4+35248\tau^3-122724\tau^2+195220\tau-110355)}}$,\\
\vskip .2cm
$c^{(\zeta=2)}_{4,2}(\tau,8)=-\sqrt{5184(2\tau-1)(\tau-5)(\tau-6)(\tau-7)\over{11(4\tau-13)(256\tau^5-4800\tau^4+35248\tau^3-122724\tau^2+195220\tau-110355)}}$,\\
\vskip .2cm
$c^{(\zeta=2)}_{5,0}(\tau,8)=0$,~
$c^{(\zeta=2)}_{5,2}(\tau,8)=\sqrt{512(2\tau-1)(\tau-4)(\tau-5)(\tau-6)(\tau-7)\over{(4\tau-13)(256\tau^5-4800\tau^4+35248\tau^3-122724\tau^2+195220\tau-110355)}}$,\\
\vskip .2cm
$c^{(\zeta=2)}_{6,2}(\tau,8)=-\sqrt{2048(2\tau-1)(\tau-3)(\tau-4)(\tau-5)(\tau-6)(\tau-7)\over{7(4\tau-13)(256\tau^5-4800\tau^4+35248\tau^3-122724\tau^2+195220\tau-110355)}}$,\\
\vskip .2cm
$c^{(\zeta=2)}_{8,4}(\tau,8)=\sqrt{1024(\tau-2)(\tau-3)(\tau-4)(\tau-5)(\tau-6)(\tau-7)\over{5(4\tau-13)(256\tau^5-4800\tau^4+35248\tau^3-122724\tau^2+195220\tau-110355)}}$.
\end{center}
After the Gram-Schmidt orthonormalization, we have

\begin{center}
$\bar{\bf c}^{(\chi=1)}(\tau,8)={\bf c}^{(\zeta=1)}(\tau,8)$;
\vskip .2cm
$\bar{c}^{(\chi=2)}_{0,0}(\tau,8)=\sqrt{30240(3\tau-7)^2(2\tau+1)(2\tau-1)(\tau-5)(\tau-6)(\tau-7)\over{
(4\tau-13)(4\tau-11)(4\tau-9)(4\tau-7)(2\tau-5)(2\tau-7)(\tau-3)[4\tau(2\tau-13)(4\tau^2-26\tau+97)+4515]}}$,\\
\vskip .2cm
$\bar{c}^{(\chi=2)}_{1,0}(\tau,8)=-\sqrt{23040(3\tau-7)^2(\tau-7)^2(2\tau+1)(2\tau-1)(\tau-5)(\tau-6)\over{
(4\tau-13)(4\tau-11)(4\tau-9)(4\tau-7)(2\tau-5)(2\tau-7)(\tau-3)[4\tau(2\tau-13)(4\tau^2-26\tau+97)+4515]}}$,\\
\vskip .2cm
$\bar{c}^{(\chi=2)}_{2,0}(\tau,8)=\sqrt{92160(3\tau-7)^2(\tau-6)^2(\tau-7)^2(2\tau+1)(2\tau-1)(\tau-5)\over{11
(4\tau-13)(4\tau-11)(4\tau-9)(4\tau-7)(2\tau-5)(2\tau-7)(\tau-3)[4\tau(2\tau-13)(4\tau^2-26\tau+97)+4515]}}$,\\
\vskip .2cm
$\bar{c}^{(\chi=2)}_{3,0}(\tau,8)=-\sqrt{(128\tau^4-2624\tau^3+19312\tau^2-59716\tau+63735)^2(2\tau+1)(2\tau-1)\over{{
(4\tau-13)(4\tau-11)(4\tau-9)(4\tau-7)(2\tau-5)(2\tau-7)(\tau-3)[4\tau(2\tau-13)(4\tau^2-26\tau+97)+4515]}}}$,\\
\vskip .2cm
$\bar{c}^{(\chi=2)}_{4,0}(\tau,8)=\sqrt{(256\tau^4-5568\tau^3+42224\tau^2-132612\tau+142695)^2(2\tau+1)(2\tau-1)(\tau-4)\over{35{
(4\tau-13)(4\tau-11)(4\tau-9)(4\tau-7)(2\tau-5)(2\tau-7)(\tau-3)[4\tau(2\tau-13)(4\tau^2-26\tau+97)+4515]}}}$,\\
\vskip .2cm
$\bar{c}^{(\chi=2)}_{4,2}(\tau,8)=-\sqrt{162[4\tau(2\tau-13)(4\tau^2-26\tau+97)+4515](2\tau-1)\over{11{
(4\tau-13)(4\tau-11)(4\tau-9)(4\tau-7)(2\tau-5)(2\tau-7)(\tau-3)}}}$,\\
\vskip .2cm
$\bar{c}^{(\chi=2)}_{5,0}(\tau,8)=\sqrt{4(64\tau^3-720\tau^2+2636\tau-3045)^2(2\tau+1)(2\tau-1)(\tau-4)\over{{
(4\tau-13)(4\tau-11)(4\tau-9)(4\tau-7)(2\tau-5)(2\tau-7)[4\tau(2\tau-13)(4\tau^2-26\tau+97)+4515]}}}$,\\
\vskip .2cm
$\bar{c}^{(\chi=2)}_{5,2}(\tau,8)=\sqrt{16[4\tau(2\tau-13)(4\tau^2-26\tau+97)+4515](2\tau-1)(\tau-4)\over{{
(4\tau-13)(4\tau-11)(4\tau-9)(4\tau-7)(2\tau-5)(2\tau-7)(\tau-3)}}}$,\\
\vskip .2cm
$\bar{c}^{(\chi=2)}_{6,2}(\tau,8)=-\sqrt{64(32\tau^4-416\tau^3+2128\tau^2-5044\tau+4515)^2(2\tau-1)(\tau-4)\over{7{
(4\tau-13)(4\tau-11)(4\tau-9)(4\tau-7)(2\tau-5)(2\tau-7)[4\tau(2\tau-13)(4\tau^2-26\tau+97)+4515]}}}$,\\
\vskip .2cm
$\bar{c}^{(\chi=2)}_{8,4}(\tau,8)=\sqrt{32(32\tau^4-416\tau^3+2128\tau^2-5044\tau+4515)^2(\tau-2)(\tau-4)\over{5{
(4\tau-13)(4\tau-11)(4\tau-9)(4\tau-7)(2\tau-5)(2\tau-7)[4\tau(2\tau-13)(4\tau^2-26\tau+97)+4515]}}}$.
\end{center}

When $4\leq\tau\leq 7$, similar to discussions in previous examples,
the expansion coefficients $c_{0,0}(\tau,8)$, $c_{1,0}(\tau,8)$,  and $c_{2,0}(\tau,8)$  become zero
for  $L=2\tau-8$. The effective projection matrix ${\bf P}(\tau,8)$
in this case is reduced as

$${\bf P}(\tau,8)=\left(\begin{matrix}
6\sqrt{2}&0 &\sqrt{4(2\tau+1)\over{99}} &0 &0 &0&0\cr
\sqrt{14(2\tau-8)\over{3}} &\sqrt{60}&0 &0 &\sqrt{4(2\tau+1)\over{21}}  &0&0\cr
0 &0 &0 &0 &0 &0 &0\cr
0 &\sqrt{12(2\tau-6)\over{7}} &0 &\sqrt{30} &0 &\sqrt{24(2\tau+1)\over{35}}&0 \cr
0&0 &\sqrt{88(2\tau-8)\over{9}}&0 &3\sqrt{2}&0 &0\cr
0 &0 &0 &0 &\sqrt{36(2\tau-6)\over{7}} &3\sqrt{2} &0\cr
0&0 &0 &0&0 &\sqrt{14(2\tau-4)\over{15}} &\sqrt{8(2\tau-1)\over{3}}\cr
\end{matrix}\right)\eqno(50)$$
with the remaining nonzero components of ${\bf c}^{T}(\tau,8)$ arranged as
$\left( c_{3,0}(\tau,8),c_{4,0}(\tau,8),c_{4,2}(\tau,8), c_{5,0}(\tau,8),c_{5,2}(\tau,8)\right.$,
$\left. c_{6,2}(\tau,8),c_{8,4}(\tau,8)\right)$.
It can be shown that no nonzero solution of ${\bf c}(\tau,8)$ exists from (50) when $\tau=4$.
For $5\leq\tau\leq 7$, $L=2\tau-8$ occurs only once with the orthonormalized expansion coefficients

\begin{center}
$c_{3,0}(\tau,8)=\sqrt{(2\tau+1)(2\tau-1)\over{{
(4\tau-13)(36\tau^3+620\tau^2-2517\tau+2023)}}}$,~
${c}_{4,0}(\tau,8)=-\sqrt{289(2\tau+1)(2\tau-1)(\tau-4)\over{{
35(4\tau-13)(36\tau^3+620\tau^2-2517\tau+2023)}}}$,\\
\vskip .2cm
${c}_{4,2}(\tau,8)=-\sqrt{1782(2\tau-1)\over{{
(4\tau-13)(36\tau^3+620\tau^2-2517\tau+2023)}}}$,~
${c}_{5,0}(\tau,8)=-\sqrt{36(2\tau+1)(2\tau-1)(\tau-3)(\tau-4)\over{{
(4\tau-13)(36\tau^3+620\tau^2-2517\tau+2023)}}}$,\\
\vskip .2cm
${c}_{5,2}(\tau,8)=\sqrt{1936(2\tau-1)(\tau-4)\over{{
(4\tau-13)(36\tau^3+620\tau^2-2517\tau+2023)}}}$,~
$c_{6,2}(\tau,8)=-\sqrt{7744(2\tau-1)(\tau-3)(\tau-4)\over{{
7(4\tau-13)(36\tau^3+620\tau^2-2517\tau+2023)}}}$,\\
\vskip .2cm
${c}_{8,4}(\tau,8)=\sqrt{3872(\tau-2)(\tau-3)(\tau-4)\over{{
5(4\tau-13)(36\tau^3+620\tau^2-2517\tau+2023)}}}$.~~
~~~~~~~~~~~~~~~~~~~~~~~~~~~~~~~~~~~~~~~~~~~~~~~~~~~~\\
\end{center}

As shown from the above examples, it seems that
the orthonormalized expansion coefficients $\bar{{\bf c}}^{(\chi)}(\tau,k)$ can always be expressed
by polynomials of $\tau$. But the expression becomes much more complicated
with increasing of $k$, especially for non-multiplicity-free cases.
Anyway, $\lambda=0$ solutions of (41) determined by (39) for given $\tau$ and $k$
completely determine the expansion coefficients $\bar{{\bf c}}^{(\chi)}(\tau,k)$,
of which a numerical algorithm can easily be implemented for the
purpose. The results are consistent with the multiplicities calculated
from the well-known  $O(5)\downarrow O(3)$ branching rule for symmetric irrep $(\tau~0)$ of $O(5)$
shown in \cite{will,corr,kem}
with $L=2p,~2p-2,2p-3,\cdots,p$ and $p=\tau,\tau-3,\tau-6,\cdots,p_{\min}$,
where $p_{\min}=0,1,2$.

Moreover, as shown in \cite{pandra}, there is
an arbitrary $SO(\rm{Multi}(\tau,k))$ rotational transformation with respect to  the multiplicity
labels $\chi=1,2,\cdots,\rm{Multi}(\tau,k)$.  When $\rm{Multi}(\tau,k)=2$
for example, let
$\vert\chi=1\rangle=\left\vert
\begin{array}{l}
~~~~~~~~~~~(\tau~0)\\
\chi=1, L=M_{L}=2\tau-k\\
\end{array}\right\rangle$
and $\vert\chi=2\rangle=\left\vert
\begin{array}{l}
~~~~~~~~~~~(\tau~0)\\
\chi=2, L=M_{L}=2\tau-k\\
\end{array}\right\rangle$
be orthonormalized basis vectors of $O(5)\supset O(3)$.
New vectors $\{\vert\bar{\chi}\rangle\}$
after an $SO(2)$ rotation with respect to the
multiplicity labels with

$$\vert\bar{\chi}=1\rangle=\cos\theta\vert\chi=1\rangle-\sin\theta\vert\chi=2\rangle,$$
$$\vert\bar{\chi}=2\rangle=\sin\theta\vert\chi=1\rangle+\cos\theta\vert\chi=2\rangle\eqno(51)$$
are also orthonormalized
basis vectors of $O(5)\supset O(3)$, where $0\leq\theta\leq 2\pi$.
As a result,  non-multiplicity-free Wigner coefficients  of $O(5)\supset O(3)$
may be numerically different when they are derived by using different methods.

\vskip .3cm
\section{Some elementary Wigner coefficients of $O(5)\supset O(3)$}

Once the expansion coefficients $\bar{{\bf c}}^{(\chi)}(\tau,k)$ are obtained,
one can easily calculate  matrix elements of $d$-boson creation operators $\{b^{\dagger}_{-2}, b^{\dagger}_{-1},
\cdots,b^{\dagger}_{2}\}$ in the $O(5)\supset O(3)$ basis.
Since $\{b^{\dagger}_{-2}, b^{\dagger}_{-1},
\cdots,b^{\dagger}_{2}\}$ is rank-$1$ and rank-$2$ irreducible tensor operators of
$O(5)$ and $O(3)$, respectively,  using the Racah factorization lemma, we have

$$\left\langle
\begin{array}{l}
~~~~(\tau+1~ 0)\\
\chi^{\prime}, L= M_{L}=2\tau+2-k^{\prime}\\
\end{array}\right\vert b^{\dagger}_{\mu}\left\vert
\begin{array}{l}
~~~~~(\tau~ 0)\\
\chi, L= M_{L}=2\tau-k\\
\end{array}\right\rangle=\sqrt{\tau+1}\left\langle\begin{array}{c}
(\tau~0)\cr
\chi,~2\tau-k\cr
\end{array}
\begin{array}{c}
(1~0)\cr
~2\cr
\end{array}\right\vert\left.
\begin{array}{c}
(\tau+1~0)\cr
\chi^{\prime},~2\tau+2-k^{\prime}\cr
\end{array}
\right\rangle\times
$$
$$\left\langle 2\tau-k,~2\tau-k;~2~\mu\vert 2\tau+2-k^{\prime},~2\tau+2-k^{\prime}\right\rangle,\eqno(52)$$
where the condition $k^{\prime}=k+2-\mu$ should be satisfied to keep the $O(3)$ CG coefficient
$\left\langle 2\tau-k,~2\tau-k;~2~\mu\vert 2\tau+2-k^{\prime},~2\tau+2-k^{\prime}\right\rangle$ nonzero
in order to derive the corresponding elementary
$O(5)\supset O(3)$ Wigner coefficient $\left\langle\begin{array}{c}
(\tau~0)\cr
\chi,~2\tau-k\cr
\end{array}
\begin{array}{c}
(1~0)\cr
2\cr
\end{array}\right\vert\left.
\begin{array}{c}
(\tau+1~0)\cr
\chi^{\prime},~2\tau+2-k^{\prime}\cr
\end{array}
\right\rangle$.
After the left hand side of Eq. (52) is expanded in terms of $O(5)\supset O_{1}(3)\otimes U(1)$ basis vectors
according to (37) with orthonormalized expansion coefficients $\bar{{\bf c}}^{(\chi)}(\tau)$,  we obtain

$$\left\langle\begin{array}{c}
(\tau~0)\cr
\chi,~2\tau-k\cr
\end{array}
\begin{array}{c}
(1~0)\cr
~2\cr
\end{array}\right\vert\left.
\begin{array}{c}
(\tau+1~0)\cr
\chi^{\prime},~2\tau-k+\mu\cr
\end{array}
\right\rangle=
\sqrt{(4\tau+\mu-2k+3)!(4\tau+\mu-2k-2)!\over{(4\tau+2\mu-2k)!(4\tau-2k)!(4\tau+2\mu-2k+1)(\tau+1)}}\times
$$
$$\sum_{q^{\prime}t^{\prime}qt}\bar{c}_{q^{\prime}t^{\prime}}^{(\chi^{\prime})}
(\tau+1,k+2-\mu)\bar{c}^{(\chi)}_{qt}(\tau,k)\times$$
$$
\left\langle
\begin{array}{l}
~~~~~~~~~~~~~~~~~~~~~~~~~~~~~~~~~~~~~~(\tau+1~0)\\
k+2-\mu-q^{\prime}~k+2-\mu-2q^{\prime}+2t^{\prime}, {1\over{2}}(\tau-1-k+\mu+q^{\prime}-t^{\prime})\\
\end{array}
\right\vert b^{\dagger}_{\mu}\left\vert
\begin{array}{l}
~~~~~~~~~~~~~~~~~~~~~~~(\tau~0)\\
k-q,~k-2q+2t,~{1\over{2}}(\tau-k+q-t)\\
\end{array}\right\rangle\eqno(53)
$$
for $\mu=2,1,0,-1,-2$,
where the matrix elements of $d$-boson operators under the $O(5)\supset O_{1}(3)\otimes U(1)$ basis
in the sum are all given in Sec. III.
For the specific values of $\mu$, (53) can be simplified with

$$\left\langle\begin{array}{c}
(\tau~0)\cr
\chi,~2\tau-k\cr
\end{array}
\begin{array}{c}
(1~0)\cr
~2\cr
\end{array}\right\vert\left.
\begin{array}{c}
(\tau+1~0)\cr
\chi^{\prime},~2\tau+2-k\cr
\end{array}
\right\rangle=
\sum_{qt}\bar{c}_{qt}^{(\chi^{\prime})}(\tau+1,k)\bar{c}^{(\chi)}_{qt}(\tau,k)
\sqrt{(2\tau+3-t)(2\tau-2k+2q-t+2)\over{2(\tau+1)(2\tau+3)}},\eqno(54)
$$

$$\left\langle\begin{array}{c}
(\tau~0)\cr
\chi,~2\tau-k\cr
\end{array}
\begin{array}{c}
(1~0)\cr
~2\cr
\end{array}\right\vert\left.
\begin{array}{c}
(\tau+1~0)\cr
\chi^{\prime},~2\tau+1-k\cr
\end{array}
\right\rangle=
\sum_{qt}\bar{c}_{qt}^{(\chi^{\prime})}(\tau+1,k+1)\bar{c}^{(\chi)}_{qt}(\tau,k)\times$$

$$\sqrt{(2\tau-k+2)(2\tau+3-t)(2k-2q+t+3)(2k-3q+2t+2)(2k-3q+2t+1)
\over{2(\tau+1)(2\tau+3)(2\tau-k)(2k-2q+1)(2k-2q+3)}}+$$

$$\sum_{qt}\bar{c}_{q+2~t+2}^{(\chi^{\prime})}(\tau+1,k+1)\bar{c}^{(\chi)}_{qt}(\tau,k)
\sqrt{(2\tau-k+2)(2\tau-2k+2q-t)(t+2)(q-2t)(q-2t-2)\over{2(\tau+1)(2\tau+3)(2\tau-k)(2k-2q+1)(2k-2q-1)}},\eqno(55)
$$

$$\left\langle\begin{array}{c}
(\tau~0)\cr
\chi,~2\tau-k\cr
\end{array}
\begin{array}{c}
(1~0)\cr
~2\cr
\end{array}\right\vert\left.
\begin{array}{c}
(\tau+1~0)\cr
\chi^{\prime},~2\tau-k\cr
\end{array}
\right\rangle=
\sum_{qt}\bar{c}_{q+1~t}^{(\chi^{\prime})}(\tau+1,k+2)\bar{c}^{(\chi)}_{qt}(\tau,k)\times$$

$$\sqrt{(4\tau-2k+3)(2\tau+3-t)(2\tau-k+1)(2k-2q+t+3)(2k-3q+2t+1)(q-2t+1)
\over{(\tau+1)(2\tau+3)(2\tau-k)(4\tau-2k-1)(2k-2q+1)(2k-2q+3)}}+
\sum_{qt}\bar{c}_{q+3~t+2}^{(\chi^{\prime})}(\tau+1,k+2)\times$$
$$\bar{c}^{(\chi)}_{qt}(\tau,k)
\sqrt{(4\tau-2k+3)(2\tau-k+1)(2\tau-2k+2q-t+2)(t+2)(2k-3q+2t)(q-2t)
\over{(\tau+1)(2\tau+3)(4\tau-2k-1)(2\tau-k)(2k-2q+1)(2k-2q-1)}},\eqno(56)
$$

$$\left\langle\begin{array}{c}
(\tau~0)\cr
\chi,~2\tau-k\cr
\end{array}
\begin{array}{c}
(1~0)\cr
~2\cr
\end{array}\right\vert\left.
\begin{array}{c}
(\tau+1~0)\cr
\chi^{\prime},~2\tau-k-1\cr
\end{array}
\right\rangle=
\sum_{qt}\bar{c}_{q+2~t}^{(\chi^{\prime})}(\tau+1,k+3)\bar{c}^{(\chi)}_{qt}(\tau,k)\times$$

$$\sqrt{(4\tau-2k+1)(2\tau+3-t)(2\tau-k+1)(2k-2q+t+3)(2k-3q+2t+2)(q-2t+1)
\over{2(\tau+1)(2\tau+3)(2\tau-k-1)(4\tau-2k-1)(2k-2q+1)(2k-2q+3)}}
+\sum_{qt}\bar{c}_{q+4~t+2}^{(\chi^{\prime})}(\tau+1,k+3)\times$$

$$\bar{c}^{(\chi)}_{qt}(\tau,k)
\sqrt{(4\tau-2k+1)(2\tau-k+1)(2\tau-2k+2q-t+2)(t+2)(2k-3q+2t)(2k-3q+2t-1)
\over{2(\tau+1)(2\tau+3)(4\tau-2k-1)(2\tau-k-1)(2k-2q+1)(2k-2q-1)}},\eqno(57)
$$

$$\left\langle\begin{array}{c}
(\tau~0)\cr
\chi,~2\tau-k\cr
\end{array}
\begin{array}{c}
(1~0)\cr
~2\cr
\end{array}\right\vert\left.
\begin{array}{c}
(\tau+1~0)\cr
\chi^{\prime},~2\tau-k-2\cr
\end{array}
\right\rangle=
\sum_{qt}\bar{c}_{q+4~t+2}^{(\chi^{\prime})}(\tau+1,k+4)\bar{c}^{(\chi)}_{qt}(\tau,k)\sqrt{(4\tau-2k+1)(4\tau-2k+1)(t+2)
\over{2(\tau+1)(2\tau+3)(4\tau-2k-3)}}.\eqno(58)
$$

For multiplicity-free cases, our results are consistent with
those derived in \cite{rt} up to a phase. Let
$\left\langle\begin{array}{c}
(\tau~0)\cr
~L_{1}\cr
\end{array}
\begin{array}{c}
(1~0)\cr
2\cr
\end{array}\right\vert\left.
\begin{array}{c}
(\tau+1~0)\cr
L\cr
\end{array}
\right\rangle_{\rm R}$
be  multiplicity-free  Wigner coefficients of $O(5)\supset O(3)$ obtained numerically in \cite{rt}. The
$O(5)\supset O(3)$ Wigner coefficients derived from (54)-(58) can be expressed as

$$\left\langle\begin{array}{c}
(\tau~0)\cr
~L_{1}\cr
\end{array}
\begin{array}{c}
(1~0)\cr
2\cr
\end{array}\right\vert\left.
\begin{array}{c}
(\tau+1~0)\cr
L\cr
\end{array}
\right\rangle=(-)^{L_{1}+2-L}\left\langle\begin{array}{c}
(\tau~0)\cr
~L_{1}\cr
\end{array}
\begin{array}{c}
(1~0)\cr
2\cr
\end{array}\right\vert\left.
\begin{array}{c}
(\tau+1~0)\cr
L\cr
\end{array}
\right\rangle_{\rm R}.\eqno(59)$$
While non-multiplicity-free  Wigner coefficients of $O(5)\supset O(3)$
derived from (54)-(58) are numerically different as compared to the corresponding
numerical results shown in \cite{rt}. But they all satisfy the orthonormality
condition:

$$\sum_{\chi_{1}L_{1}}\left\vert\left\langle\begin{array}{c}
~~~(\tau~0)\cr
\chi_{1} ~L_{1}\cr
\end{array}
\begin{array}{c}
(1~0)\cr
2\cr
\end{array}\right\vert\left.
\begin{array}{c}
~~~(\tau+1~0)\cr
\chi~L\cr
\end{array}
\right\rangle\right\vert^2=1.\eqno(60)$$

\begin{table}[htb]
\caption[c]{Elementary $O(5)\supset O(3)$ Wigner coefficients
$\left\langle\begin{array}{c}
(\tau~0)\cr
~L_{1}\cr
\end{array}
\begin{array}{c}
(1~0)\cr
2\cr
\end{array}\right\vert\left.
\begin{array}{c}
(\tau+1~0)\cr
L\cr
\end{array}
\right\rangle$.
}
\begin{tabular}{cccccc}
\hline
$L_1$~~&$L=2\tau-1$~~~&$L=2\tau$~~~ &$L=2\tau+2$\\
\hline\\
$2\tau$~~&$\sqrt{2(4\tau+1)(\tau-1)\over{(4\tau-1)(2\tau-1)(\tau+1)}}$~~~ &$-\sqrt{2(2\tau+1)\over{(4\tau-1)(\tau+1)}}$
&$1$\\\\
$2\tau-2$~~&$\sqrt{3(2\tau+1)\over{(4\tau-1)(\tau+1)(\tau-1)}}$~~~ &$\sqrt{(\tau-1)(4\tau+3)\over{(4\tau-1)(\tau+1)}}$&$0$\\\\
$2\tau-3$~~&$\sqrt{\tau(2\tau+1)(\tau-2)\over{(2\tau-1)(\tau+1)(\tau-1)}}$&$0$&$0$\\
\hline\hline

$L_1$~~&$L=2\tau-2$&$L=2\tau-3$ \\
\hline\\
$2\tau$~~&$-\sqrt{32\tau(\tau-1)(\tau-2)\over{(4\tau-1)(4\tau-3)(2\tau+3)(2\tau-1)(\tau+1)}}$ &$0$
\\\\
$2\tau-2$~~&$-\sqrt{4(2\tau+1)^2(4\tau+1)(\tau-2)\over{(4\tau-1)(4\tau-5)(2\tau+3)(\tau+1)(\tau-1)}}$
&$\sqrt{2(4\tau-1)(\tau-2)(\tau-3)\over{(4\tau-5)(2\tau-3)(\tau+1)(\tau-1)}}$\\\\
$2\tau-3$~~&$\sqrt{4(4\tau+1)(4\tau-1)\over{(2\tau+3)(2\tau-3)(2\tau-1)(\tau+1)(\tau-1)}}$
&$-\sqrt{2(2\tau+1)^2(\tau-3)\over{(4\tau-7)(2\tau-3)(\tau+1)(\tau-1)}}$\\\\
$2\tau-4$~~&$\sqrt{(\tau-3)(2\tau-1)(2\tau+1)(4\tau+1)(4\tau-1)\over{(4\tau-3)(4\tau-5)(2\tau+3)(2\tau-3)(\tau+1)}}$
&$\sqrt{6(2\tau+1)(2\tau-1)\over{(4\tau-5)(2\tau-3)(\tau+1)(\tau-2)}}$\\\\
$2\tau-5$~~&$0$&$\sqrt{(\tau-1)(\tau-4)(4\tau-3)(2\tau-1)\over{(4\tau-7)(2\tau-3)(\tau+1)(\tau-2)}}$\\
\hline
\hline
\end{tabular}
\end{table}

\begin{table}[htb]
\caption[c]{Elementary $O(5)\supset O(3)$ Wigner coefficients
$\left\langle
\begin{array}{c}
~~(\tau~0)\cr
\chi_{1}~L_{1}\cr
\end{array}
\begin{array}{c}
(1~0)\cr
2\cr
\end{array}\right\vert\left.
\begin{array}{c}
(\tau+1~0)\cr
\chi~L=2\tau-4\cr
\end{array}
\right\rangle$.
}
\begin{tabular}{cccccc}
\hline
$\chi_{1}$,~$\L_1$~~&$\chi=1$&$\chi=2$ \\
\hline\\
$2\tau-2$~~&$0$
&$\sqrt{3f(\tau+1)\over{(4\tau-7)(4\tau-5)(2\tau+3)
(2\tau-3)(\tau+1)(\tau-1)}}$\\\\
$2\tau-3$~~&$\sqrt{32(4\tau-5)(2\tau-1)(\tau-1)^2(\tau-3)(\tau-4)\over{
(4\tau-7)(\tau+1)(\tau-2)f(\tau+1)}}$
&$\sqrt{(4\tau-9)^2(4\tau-3)(4\tau-1)(2\tau+1)^2(2\tau-1)\over{(4\tau-7)(2\tau+3)
(2\tau-3)(\tau+1)(\tau-1)f(\tau+1)}}$\\\\
$2\tau-4$~~&$-\sqrt{6(4\tau-5)^2(4\tau-3)(2\tau+1)(\tau-4)\over{
(4\tau-9)(\tau+1)(\tau-2)f(\tau+1)}}$
&$\sqrt{192(4\tau-1)(2\tau+1)(2\tau-1)^2(\tau-1)(\tau-3)^3\over{(4\tau-9)(4\tau-5)
(2\tau+3)(2\tau-3)(\tau+1)f(\tau+1)}}$\\\\
$2\tau-5$~~&$\sqrt{12(2\tau+1)^2(2\tau-1)^2(2\tau-3)^2\over{
(4\tau-7)(2\tau-5)(\tau+1)(\tau-2)f(\tau+1)}}$
&$-\sqrt{96(4\tau-5)(4\tau-3)(4\tau-1)(\tau-1)(\tau-3)(\tau-4)\over{(4\tau-7)(2\tau+3)
(2\tau-3)(2\tau-5)(\tau+1)f(\tau+1)}}$\\\\
$\chi_{1}=1,~2\tau-6$~~&$\sqrt{(\tau-5)f(\tau+1)\over{(\tau+1)f(\tau)}}$
&$0$\\\\
$\chi_{1}=2,~2\tau-6$~~&$\sqrt{32(4\tau-7)(4\tau-5)(2\tau+1)(2\tau-1)^2(2\tau-3)^2(\tau-3)(\tau-4)
\over{(4\tau-9)(2\tau-5)(\tau+1)(\tau-2)f(\tau+1)f(\tau)}}$
&$\sqrt{(4\tau-3)(4\tau-1)(2\tau+1)(2\tau-3)(\tau-1)f(\tau)\over{(4\tau-9)(4\tau-7)(2\tau+3)
(2\tau-5)(\tau+1)f(\tau+1)}}$\\
\hline\hline
\end{tabular}
{$f(\tau)=32\tau^4-288\tau^3+1024\tau^2-1692\tau+1065.~
~~~~~~~~~~~~~~~~~~~~~~~~~~~~~~~~~~~~~~~~~~~~~~~~~~~~~~~~~~~~~~~~$}
\end{table}

As discussed at the end of previous section, though
non-multiplicity-free  Wigner coefficients
derived from different methods may be different in values,
they are equivalent up to an $SO({\rm Multi}(\tau,k))$ rotational
transformation. Furthermore,
similar to the symmetry property of $O(5)\supset O_1(3)\otimes U(1)$ discussed
in Sec. III,
the $O(5)\supset O(3)$ Wigner coefficients
satisfy the following symmetry relations as discussed in many papers, for example in \cite{sun,rt}:

$$\left\langle\begin{array}{c}
(\tau~0)\cr
\chi_{1}~L_{1}\cr
\end{array}
\begin{array}{c}
(1~0)\cr
2\cr
\end{array}\right\vert\left.
\begin{array}{c}
(\tau+1~0)\cr
~\chi~L\cr
\end{array}
\right\rangle=(-)^{L_{1}+2-L}\left\langle\begin{array}{c}
(1~0)\cr
2\cr
\end{array}
\begin{array}{c}
(\tau~0)\cr
\chi_{1}~L_{1}\cr
\end{array}\right\vert\left.
\begin{array}{c}
(\tau+1~0)\cr
~\chi~L\cr
\end{array}
\right\rangle\eqno(61)$$
and
$$\left\langle\begin{array}{c}
(\tau+1~0)\cr
~\chi~L\cr
\end{array}
\begin{array}{c}
(1~0)\cr
2\cr
\end{array}\right\vert\left.
\begin{array}{c}
(\tau~0)\cr
\chi_{1}~L_{1}\cr
\end{array}
\right\rangle=(-)^{L_{1}+2-L}\sqrt{\dim(\tau~0)(2L+1)\over{\dim(\tau+1 ~0)(2L_1+1)}}
\left\langle\begin{array}{c}
(\tau~0)\cr
\chi_{1}~L_{1}\cr
\end{array}
\begin{array}{c}
(1~0)\cr
2\cr
\end{array}\right\vert\left.
\begin{array}{c}
(\tau+1~0)\cr
~\chi~L\cr
\end{array}
\right\rangle.\eqno(62)$$

Some analytical expressions of
elementary $O(5)\supset O(3)$ Wigner coefficients for the coupling
$(\tau~0)\otimes(1~0)$ with resultant $O(3)$ quantum number $L=2\tau+2-k$
and $k\leq 6$ are shown in Tables III and IV, in which only
$\tau>k_{1}$ and $\tau>k-1$  cases related with
$L_1=2\tau-k_{1}$ and $L=2\tau+2-k$, respectively, are shown.

\section{conclusion}
In this paper, a recursive method for construction of symmetric  {irreps}
of $O(2l+1)$ in  {an} $O(2l+1)\supset O(3)$ basis for identical boson systems is proposed.
The formalism is realized based on the group chain $U(2l+1)\supset U(2l-1)\otimes U(2)$,
 {for} which the symmetric  {irreps} are simply reducible.  {Within this framework,}
the basis vectors of the $O(2l+1)\supset O(2l-1)\otimes U(1)$  {are constructed}
from those of $U(2l+1)\supset  U(2l-1)\otimes U(2)\supset O(2l-1)\otimes U(1)$
with no boson pairs,  {and from these one
can deduce} symmetric  {irreps}  of $O(2l+1)$ in the $O(2l-1)\otimes U(1)$
basis when all symmetric irreps of $O(2l-1)$ are known.

As a starting point, basis vectors of
symmetric irreps of $O(5)$ are constructed in the $O_{1}(3)\otimes U(1)$ basis.
Matrix representations of  $O(5)\supset O_{1}(3)\otimes U(1)$, together with
the elementary Wigner coefficients, are  {then generated, and after}
angular momentum projection, a three-term relation  {for} determining the expansion
coefficients of the $O(5)\supset O(3)$ basis vectors expanded in terms of those of
the $O_{1}(3)\otimes U(1)$ is derived. The eigenvectors  {with zero eigenvalues}
of the projection matrix constructed according to the three-term relation
completely determine the  basis vectors of
$O(5)\supset O(3)$, which enables  {one} to derive analytical expressions
of elementary Wigner coefficients of $O(5)\supset O(3)$ with the formulae
shown  {in (54)-(58). Some simple}  elementary Wigner coefficients of $O(5)\supset O(3)$
 {are presented as examples.
An algorithm that satisfies the three-term relation (39)
can be readily determined.} As far as the
elementary Wigner coefficients of $O(5)\supset O(3)$  {are} concerned,
the procedure shown in this paper seems simpler than the
method shown in \cite{rt} using the overlap integrals
of $O(5)$ spherical harmonic functions and the recursive
method proposed in~\cite{sun}.

Using the matrix representations
of $O(5)\supset O_{1}(3)\otimes U(1)$  { as determined above, one can
construct matrix representations of $O(7)\supset O(5)\otimes U_{3}(1)$
in a similar way,} where the generator of $U_{3}(1)$ is
${1\over{2}}(b_{3}^{\dagger}b_{3}-b_{-3}^{\dagger}b_{-3})$, with which
one can construct $O(7)\supset O(3)$ basis vectors from those
of  $O(7)\supset O(5)\otimes U_{3}(1)$.
 {A similar} procedure for identical fermion systems is also possible
because anti-symmetric irreps of $Sp(2j+1)$
in the reduction $Sp(2j+1)\downarrow Sp(2j-1)\otimes U(1)$
are also simply  {reducible, and this in turn suggests that one can establish
a similar recursive procedure to construct basis} vectors of
$Sp(2j+1)\supset O(3)$ in terms of those of $Sp(2j+1)\downarrow Sp(2j-1)\otimes U(1)$.
The related work is in progress.

\begin{acknowledgments}
One of the authors (PF) is grateful to Faculty of Science, The University of
Queensland, for support through an Ethel Raybould Visiting Fellowship.
Support from U.S. National Science Foundation (OCI-0904874),
Southeastern Universities Research Association,
Natural Science Foundation of China (11175078), Australian Research Council
(DP110103434), Doctoral Program Foundation of the State Education Ministry
of China (20102136110002), and LSU--LNNU joint research program (9961) is acknowledged.
\end{acknowledgments}

\end{document}